\documentstyle[aps]{revtex}

\begin{document}
% \draft command makes pacs numbers print
\draft
\title{Quark and gluon distributions at the earliest
        stage of heavy ion collision}
\author{ A.  Makhlin}
\address{Department of Physics and Astronomy, Wayne State University,
Detroit, MI 48202}
\date{\today}
\maketitle
\begin{abstract}
Using the general framework of quantum field kinetics \cite{Mak} we consider
new principles to compute initial distribution of  quarks and gluons after the
first hard interaction of heavy ions. We start by rewriting the integral
equations of QCD in the form which is generalizations of the familiar QCD
evolution equations. These equations  describe both space-time--  and
$(x,Q^2)$--evolution before the collision, and allow one to use the $ep$ DIS
data without reference to parton  phenomenology. New technique generate
perturbation theory that avoid double count of the processes, does not contain
an artificial factorization scale, and does not require low-momentum cut-offs
since infrared behavior is  controlled by the DIS data.
\end{abstract}
% body of paper here

\bigskip
\noindent {\bf \Large 1. Introduction.}
\bigskip

Recently, there have been many calculations
\cite{Muller1,Gul1,McLer,Blaz} to find the initial
distribution of quarks and gluons in heavy ion collisions, which may
then lead to the creation of a plasma.
Contrary to other types of matter under extreme conditions, the
situation in heavy ion physics is unique, since the very existence of
quark--gluon plasma (QGP) is inseparable from the process of creation
of the  matter it consists of. Any subsequent elements of the behavior,
which can be described in classical language,
such as thermalization or a hydrodynamic regime for the QGP,
like for any other many--particle system, is possible only after
intermixing of the initial phase-space distribution.
However the process of creation of quarks and gluons is
essentially quantum by its origin.  In the present paper only this
process is considered.

Any strict formulation of a quantum--mechanical problem requires an
exact definition of two main elements: the initial state of the
system, and the observables in the expected final state.  This
reminder would  hardly be necessary if the main difficulties were not
associated with these points: On the one hand, it is unknown how
stable nuclei of the initial state are build up from quarks and
gluons; on the other, there is no clear understanding of what the
final state may look like. So even the formulation
of the problem unavoidably contains some uncertainty
which requires special care in defining these elements.
Let us begin with a discussion about what kind of problems we may
anticipate.

{\it 1. The initial state:} Ideally, as in, for example, an atomic
collision, we would define the initial state via its wave function.
The wave functions of QCD--nuclei are unknown. A natural alternative
is to use the density matrix to describe each nucleus.

Here we may assume nuclei to be well shaped objects. The uncertainty
of their boundaries does not exceed the typical Yukawa interaction
range. In the laboratory frame both nuclei are Lorentz contracted up
to a longitudinal size $R_0/\gamma \sim 0.1 fm$.  The tail of the
Yukawa potential is contracted in the same proportion.  The world lines
of the nuclei are two opposite generatrices of the light cone that has
its vertex at the interaction point.  No interaction between nuclei is
possible before they overlap geometrically. For this reason the total
density matrix of two nuclei is a direct product of the two individual
density matrices.  The spaces of states where they act do not overlap
either.

Since no exact information about  the initial state of the nuclei is
available, it seems reasonable to rely upon the following two considerations:
First, detailed information is inessential as it basically relates to the
interactions which maintain every nucleus as a QCD bound state. The energy of
the collision is incomparably higher. It is thus enough to require that density
matrix yields the given total momentum and baryonic charge as
averages of the corresponding field operators.
Second, the residual dynamical information must reveal itself in the
same way as in other inelastic processes at extreme energies,
like deep inelastic electron-proton   or muon-nucleus scattering.
This statement may appear trivial, because  structure functions
of DIS are always used for account of this information.  However, one should
keep in mind that their definition -- which does not refer to the
parton model -- is valid only for the DIS process itself.
In order to apply structure functions to other interactions, using the
parton model is considered unavoidable.

The first priority of  this study is to avoid any intermediate phenomenology.
We insist that any information taken from parallel experiment is valuable only
as long as both processes can be described by the same theory and with the same
initial data.

{\it 2. The final state:}   The final state is assumed to be some distribution
of free quarks and gluons in the perturbative vacuum. This vacuum
is considered to be the true ground state, and is free of QCD--condensates. It
is a product of the nuclear collision and is postulated to exist
in a sufficiently large volume. The spectrum of possible states
forms the continuum, which is unoccupied at the beginning of the collision.

It is supposed also that the information which is most important for
understanding the future evolution is concentrated in the single--particle
distributions of quarks and gluons. These distributions
must be calculated from their quantum--mechanical definitions, keeping in mind
how they are to be measured in a hypothetical experiment.
Two-- and more--particle distributions should be defined as
independent elements corresponding to other measurements.

In this work we rely heavily on a previous paper \cite{Mak},
where integral equations of
QCD were derived without assuming that averaging is performed
over a stationary state. To some extent these equations resemble the diagram
technique of Keldysh \cite{Keld}, which was designed for non-equilibrium
processes. They are of the same matrix form, but are not derived with a
view to obtaining quasi-classical kinetic equations. Fields and their
correlators remain the main objects of these equations and no
phase-space distributions is introduced. For this reason the approach was
named ``quantum field kinetics'' (QFK).

These new equations are the result of an initial resummation of the
perturbation series for probabilities of inclusive processes (or any other
observables). On the one hand, these
new equations create an approach which proves to be very
effective for studying various inclusive processes. On the other,
it allows one to trace the temporal evolution of a colliding system,
beginning from preparation in the past right up to the moment of
measurement. This latter feature makes the new approach extremely
attractive for our goals: The evolution of any
physical system is completely defined by
the initial data, and the act of measurement
only selects one or more of all the possible quantum trajectories.
Therefore, one can expect that results of similar types of measurement
will be similar. In this paper, we want to consider two similar
processes, {\it viz.}, deep inelastic electron-proton scattering and
``deep inelastic pp- or AA-collisions,'' in parallel.

To begin with, we must define the
observables of these processes in the same way.
Definitions of inclusive amplitudes and inclusive probabilities to find
one quark or one gluon in the final state are given in Section 2.
The DIS cross-section,
viewed as an inclusive process where nothing is measured except the
momentum of the scattered electron, is defined in the same terms at the
beginning of Section 3. The rest of this section considers an instructive
example of the lowest order calculations using a model density matrix.
This calculation allows one to introduce all spinor
and vector functions without extra complicated notations, and to clarify
the complete calculation.

The dynamical equations in their full tensor--spinor form are derived in
Section 4.  It immediately becomes clear that these equations
have a ladder structure. It appears that the well known ordering of ladder
cells by  Feynman $x$ and virtuality is a direct consequence of the
retarded temporal ordering inherent to these equations in their coordinate
form. Smaller $x$ and bigger virtualities correspond to the later times.
Thus, as a by--product, we obtain an answer on a very old question about
such correspondence \cite{Coll,Durand,Muller2}.

The new equations appear to be richer than usual QCD evolution equations
for DIS structure functions, as derived from the renormalization group
approach. The new equations interconnect two invariant functions
of the vector field, and two of the quark field.  The new evolution equations
do not depend upon the type of last interaction, however they may be projected
onto any definite process. Specific properties of
the electromagnetic interaction select only one of the spinor functions into
the definition of the structure function $F_2$ of unpolarized DIS.
However, both quark field functions remain in the evolution
equations, along with the two functions of the vector field.  The relative
scale of additional terms and their possible role is examined in Appendix~1.
It is also shown that among other extensions, the
new equations naturally include  the BFKL \cite{BFKL} equation and the
effects of quark and  gluon shadowing at small $x$. Corresponding
shadowing terms
appear to be parametrically larger than in previous derivations
\cite{GLR,MueQui}.

The meaning of the objects that obey the new evolution equation comes to light
in a discussion of renormalization. The renormalization group approach can
not be used here, if only because most of the terms in these equations
correspond to observables (imaginary part of self-energies) are finite and
may not be renormalized. Instead, we renormalize the second subgroup
of the evolution equations
for the real part of the self-energies using a conventional BHPZ scheme.
Ultraviolet divergencies are compensated for by counter-terms from the original
Lagrangian. The running coupling appears precisely from the requirement
of renormalizability. For the moment, this
part of the study is at the level of a basic
idea.

Using some structural, rather than quantitative, assumptions, and after
projecting onto specific observables of the e-p DIS,
the  new equations can be reduced to the system of GLAP
equations \cite{AP,Lip,GD}.
The objects which enter the new equations are similar to self-energies, and
we shall call them sources.
It appears that the QCD evolution proceeds in a way such as to create a source
of a field which interacts with the detector in a certain way. The evolution
causes the dynamical assembly of a special wave packet which represents
a bare quark or gluon. This process takes place in real time and ends
at the moment of interaction.

We expect DIS to provide the dynamical information about this process.
This information is valuable only as long as no measurements were done
before the last interaction.
All unobserved information (and providing it was not observed) is
included in the definition of the sources with their full dependence
upon $x$ and $Q^2$.

Here, we do not adhere  to the picture of ``wee partons,'' and do not
share the opinion that the QCD evolution equations describe how one valence
quark develops a cloud of virtual quarks and gluons around it at small
distances. We believe that they give (perturbatively) the
``evolution'' of the detector response, provided the trigger includes the
requirement  of a bare on-shell quark in the final state (the resonant
condition which results in $x_{Bj}=x_F$). That they definitely do not
correspond to the
evolution of a single quark, is seen, for example, from possibility of
including fusion.

The difference between the structure functions  and the sources can be
explained using an analogy: in condensed matter or scattering theory we
introduce two different quantities: the density of states, $~\rho(E)$,  and the
number of states below energy $~E$, $~n(E)=\int^E \rho(E) dE$.  The
structure functions then
correspond to $n(E)$, while the sources correspond to $~\rho(E)$. In
an experiment we measure  $n(E)$ (which is proportional to the allowed volume
in the phase space), rather than $\rho(E)$. From this point  of view it is not
surprising that  we eventually express observables, roughly speaking, via
$dG(x,Q^2)/dQ^2$. Different measurements study these quantities, integrating
them with  their specific ``upper limits.''  The boundary conditions for $x$--
and $Q^2$--evolution  are imposed by the measurement, rather than the
initial conditions which are controlled by momentum  conservation, sum
rules, {\it etc.} These must be set in the past  without any relation to the
($x$-$Q$)-evolution.

Cross--sections of inclusive single-quark and single-gluon production
in the lowest nonvanishing order are calculated in Section 5. All of
them contain scale--dependent and --independent terms. The latter
are much bigger than the former, and as a result the cross--sections
are expected to exhibit only weak scale dependence. The lowest order
cross--sections
are strongly peaked at low rapidities and low transverse momenta.

The next order of the perturbative expansion generated by the new evolution
equations is examined in Section 6. The new expansion does not
lead to any diagrams which duplicate those already included in the
definition of the sources (or structure functions). Any such diagrams
would carry  severe collinear singularities. We carefully examine the
infrared finiteness of the diagrams that do occur. It is
shown that they are infrared safe and that no artificial cut--offs are
necessary to find the total cross-section. Higher order perturbation terms
are not expected to present any difficulty, as their final state infrared
behavior will be shielded by the final distributions themselves.

We are led to the approach advocated in this paper in an unavoidable manner.
Ideally, one would start with a complete relativistic quantum description
of the  static proton and its interaction with the detector. Unfortunately,
the many attempts to describe the bound states of QCD (see Refs.
\cite{Shuryak} and \cite{Brodsky} for the reviews) have not yet met with real
success. To calculate the production of particles in hadronic
collisions, one is limited to reasoning along the following lines:
first, an OPE-analysis of the DIS data, which gives the structure
functions of DIS; next, a partonic interpretation of the structure
functions; and, lastly, using the factorization technique.
In the end, one still faces severe theoretical problems caused by the
soft processes, the arbitrariness of the factorization scale, {\it etc.}
Here, we try to avoid these problems ``experimentally,'' by
maximizing the use of dynamic information hidden in the DIS data.

\renewcommand{\theequation}{2.\arabic{equation}}
\setcounter{equation}{0}
\bigskip
\bigskip
\noindent {\bf \Large 2. Single-particle distributions of the partons.}
\bigskip

Let two nuclei A and B, with  momenta $P_A$  and $P_B$,  move towards
each other at almost the speed of light, and let the center-of-mass system
coincide with the laboratory frame.  We assume that the center-of-mass energy
is very large, $s>>M^2$, so that the laboratory frame is the infinite momentum
frame for both nuclei. We intend to compute one-particle distributions of
quarks and gluons created after the first interaction
of these nuclei. The inclusive
amplitudes leading to creation of one quark or one gluon from the initial
state $|in\rangle$ are  as follows,
\begin{equation}
 \langle X|d({\bf p},\sigma,i)S|in\rangle ,  \;\;\;{\rm and}\;\;\;
 \langle X|c({\bf k},\lambda,a)S|in\rangle ~ ~ ~,
\label{eq:E1}
\end{equation}
where $d^{\dag}({\bf p},\sigma,i)$ is a creation operator for an on-mass-shell
quark with momentum $p$, spin $\sigma$ and color $i$.
Similarly, the operator $c^{\dag}({\bf k},\lambda,a)$ creates an on-mass-shell
gluon with momentum $k$, polarization  $\lambda$ and color $a$.  Summing
the squared moduli of these amplitudes over a complete set of uncontrolled
states $|X\rangle$, and  averaging over the initial ensemble, we find
inclusive spectra of quarks and  gluons
\begin{equation}
{ dN_q \over d{\bf p}}
 = \sum_{\sigma,i} Sp \rho_{in}
          S^{\dag}d^{\dag}({\bf p},\sigma,i)d({\bf p},\sigma,i)S~ ~ ~,
\label{eq:E2}
\end{equation}
\begin{equation}
{ dN_g \over d{\bf k}}
 = \sum_{\lambda,a} Sp \rho_{in}
          S^{\dag}c^{\dag}({\bf k},\lambda,a)c({\bf k},\lambda,a)S~ ~ ~.
\label{eq:E3}
\end{equation}

The initial state of the colliding system consists of two
Lorentz-contracted nuclei
which are causally independent, and thus
the total density matrix is a direct product
of two independent density matrices,
\begin{equation}
\rho_{in}= \rho_{A}\otimes \rho_{B} \otimes |0_{cont}\rangle\langle0_{cont}|.
\label{eq:E4}
\end{equation}
Matrix elements of $\rho_{in}$ are obtained by sandwiching it between all
state vectors $|in\rangle$ which enter definition (\ref{eq:E1}) of inclusive
amplitude. The
density matrices $\rho_A$ and $\rho_B$  contain only bound states of
quarks and gluons in the
presence of vacuum condensates.  The latter are assumed
to be destroyed in the course of the initial
hard collision and replaced by the perturbative QCD
vacuum. Initially, all states in the continuum are unoccupied. It means that
$\rho_{in}$
contains a projector  $|0_{cont}\rangle\langle0_{cont}|$ on the vacuum
state in the continuum. So we may commute the quark Fock operators with
$S$ and $S^{\dag}$ and only commutators survive in the final result:
\begin{equation}
{ dN_q \over d{\bf p}}
 =\sum_{\sigma,i} \int d^{4}x d^{4}y \bar{\psi}^{(+)}_{p,\sigma,i}(x)
\langle{ \delta S^{\dag} \over
   \delta q_i (y) }
        {{\delta S}\over{\delta \bar{q}_i (x)}}\rangle
 \psi^{(+)}_{p,\sigma,i}(x).
\label{eq:E5}
\end{equation}
In this expression $\psi^{(+)}_{p,\sigma,i}(x) $ is the Dirac wave function
of a quark. Summing over  spin and color we get
\begin{equation}
{ dN_q \over d{\bf p}}
 =\sum_{\sigma,i} \int d^{4}x d^{4}y {e^{-ip(x-y)} \over (2\pi)^3 2p^0}
{\rm Tr} [\not p  i \Sigma_{01}^{ii}(x,y)]~,
\label{eq:E6}
\end{equation}
where the full $2\times 2$ matrix of the quark self-energy is given by
\cite{Mak}
\begin{eqnarray}
  \Sigma_{AB}(x,y)=i(-1)^{A+B}g^{2}\sum_{R,S=0}^{1} (-1)^{R+S}
   \int d \xi d \eta
  t^{a} \gamma^{\mu} {\bf G}_{AR}(x,\xi)
   \Gamma^{d,\lambda}_{RB,S}(\xi,y;\eta)
 {\bf D}^{da}_{SA,\lambda\mu}(\eta,x)~.
\label{eq:E7}
\end{eqnarray}
This formula implies that both quark and gluon correlators, ${\bf G}_{AR}$
 and ${\bf D}^{da}_{SA,\lambda\mu}$,  are averaged with the density matrix
$\rho_{in}$ given by Eq.~(\ref{eq:E4}).  In the first
approximation we may replace the exact $qqg$ vertex   by the bare one.  Then
Eq.~(\ref{eq:E6}) takes the following simple form,
\begin{equation}
p^0 { dN_q \over d{\bf p}d^4 x}
 ={g^2 \over 2(2\pi)^3} \int {d^4 k \over (2\pi)^4 }
  \{ Tr [(\not p +m) t^a \gamma^{\mu} {\bf G}^{(A)}_{01}(p-k)t^b \gamma^{\mu}
{\bf D}^{(B)ba}_{10,\nu\mu}(-k) ] + [(A) \leftrightarrow (B)] \}
\label{eq:E8}
\end{equation}
where the additional superscript $(A)$ or $(B)$ denotes that the correlation
function is averaged over the initial state of nucleus $A$ or $B$,
respectively.

A similar procedure yields the following expression
for the inclusive gluon production:
\begin{equation}
{ dN_g \over d{\bf p}}
 =\sum_{\lambda,a} \int d^{4}x d^{4}y {e^{-ip(x-y)} \over (2\pi)^3 2p^0}
 \epsilon_\mu^{(\lambda)}\epsilon_\nu^{(\lambda)}[-i\Pi_{aa}^{01,\mu\nu}(x,y)]
\label{eq:E9}
\end{equation}
where the primary definition of the gluon polarization tensor
$\Pi_{01}$ is given by
\begin{equation}
\Pi_{01,ab}^{\mu\nu}(x,y)=
i \langle{ \delta S^{\dag} \over
   \delta B_{\mu}^{b} (y) }
        {{\delta S}\over{\delta B_{\nu}^{a} (x)}}\rangle~ ~ ~.
\label{eq:E10}
\end{equation}
Its general formula was derived in \cite{Mak}:
\begin{eqnarray}
  \Pi^{\mu\nu}_{AB}(x,y)=i(-1)^{A+B}g_{r}^{2}\!\!\sum_{R,S=0}^{1}\!\!(-1)^{R+S}
[-\!\int \! d \xi d \eta \gamma^{\mu}{\bf G}_{AR}(x,\xi)
 \Gamma^{\nu}_{RS,B}(\xi,\eta;y) {\bf G}_{SA}(\eta,x) + \nonumber \\
+\int  d \xi d \eta V^{\mu\alpha\nu}_{acf}(x,\xi,\eta')
{\bf D}_{AR}^{cc',\alpha\beta}(\xi,\xi')
 {\bf V}^{\nu\beta\sigma}_{RSB;bc'f'}(\xi',\eta,y)
{\bf D}_{SA}^{f'f,\lambda\sigma}(\eta,\eta')]~~,
%%\hspace*{2cm}
\label{eq:E11}
\end{eqnarray}
and we postpone its further expansion because of the complexity of the emerging
polarization structure.  However, the main idea remains the same as
for quark production:  in first
approximation we get a product of two quark,
$ {\bf G}^{(A)}_{01}{\bf G}^{(B)}_{10}$, or two gluon,
$ {\bf D}^{(A)}_{01}{\bf D}^{(B)}_{10}$ correlation functions. Each of them
is averaged with the density matrix of only one of the two nuclei. This is in
line with the independence of the initial states of the colliding nuclei.

To continue with the straightforward calculations, we
should now specify an explicit form of the nuclear density matrix. To do
this, we should solve the confinement problem -- which is not our intention
here. Moreover, we have already argued that most of the detailed information
is unnecessary. Now we shall motivate this point by
a  ``quantum kinematic'' analysis of the extreme case when
no dynamical information is required to obtain
a qualitative understanding of the phenomenon.

Let the nuclei collide at an energy of about 100 TeV per nucleon,
when the nuclear longitudinal size  is only $10^{-4}~{\rm fm}$.
This size is much less
than any known in nuclear interactions, and one may consider the domain
where the nuclei overlap as a plane surface (or the point in the
($t,z$)-plane).
 All the subsequent dynamics takes place within the future light cone of this
point, $t^2-z^2>0,~~t>0$. As the translation invariance in $t$--
and $z$--directions is manifestly broken by the initial conditions,
one should look for the appropriate
quantum numbers (other than $p^0$ and $p^z$)
to describe the final states of the particles. The symmetry
that does survive is Lorentz invariance, and the boost defined by the operator
\begin{equation}
{\hat \nu } = t [-i{\partial  \over \partial z}]
-z [i{\partial  \over \partial t}]
= i{\partial  \over \partial \phi},
\label{eq:E12}\end{equation}
is an good quantum number. The corresponding wave functions of
the free scalar particles obey the Klein-Gordon equation,
\begin{equation}
{1\over\tau}{\partial  \over \partial \tau}
(\tau {\partial \psi \over \partial \tau}) -
{1\over\tau^2}{\partial^2\psi  \over \partial \phi^2}
+m_{\perp}^{2}\psi =0~~,
\label{eq:E13}\end{equation}
and are of the following form:
\begin{equation}
\xi_{\nu,p_\perp}(x)={e^{-\pi\nu/2}\over 2\pi 2^{3/2}}
H^{(2)}_{-i\nu}(m_{\perp}\tau) e^{-i\nu\phi}
e^{i{\vec p}_{\perp}{\vec r}_{\perp}}.
\label{eq:E14}\end{equation}
Here, we use coordinates $ (~t=\tau \cosh \phi,~z=\tau \sinh \phi,
{}~{\vec r}_{\perp})$, and denote ~$m_{\perp}^2= m^2+  p_{\perp}^2$.
The eigenfunctions $~\xi_{\nu,p_\perp}(x)$
are normalized on the space-like hypersurface $\tau=const$~ within
the future light cone of the collision point:
\begin{equation}
(\xi^{*}_{\nu,p_\perp}, \xi_{\nu',p'_\perp})
=\int \tau d\phi d^2 {\vec r}  \xi^{*}_{\nu,p_\perp}(x)
i{\stackrel{\leftrightarrow}{\partial \over \partial \tau}}
\xi_{\nu',p'_\perp}(x)
=\delta (\nu -\nu')\delta ({\vec p}_{\perp}-{\vec p'}_{\perp}).
\label{eq:E15}\end{equation}

The boost of the particle is a legitimate, but seldomly used quantum
number. It is more common to use the momentum. Let us try to find the
functions which  would behave as free plane waves, at least asymptotically.
The  wave packets with the required behavior are
\begin{eqnarray}
\Xi_{\theta,p_\perp}(x)={-i\over (2\pi)^{1/2}}\int_{-\infty}^{+\infty}
 e^{-i\nu\theta} \xi_{\nu,p_\perp}(x)
={1 \over 4\pi^{3/2}}  e^{-im_{\perp}\tau\cosh(\phi-\theta)}
e^{i{\vec p}_{\perp}{\vec r}_{\perp}}.
\label{eq:E16}
\end{eqnarray}
These wave packets represent plane waves confined to within
the future light cone
of the collision point, and can be rewritten as follows:
\begin{equation}
\Xi_{\theta,p_\perp}(x)={1 \over 2\pi 2\sqrt{\pi}}  e^{-ip^0 t+ip^z z}
e^{i{\vec p}_{\perp}{\vec r}_{\perp}},~~~p^0=m_{\perp}\cosh\theta,~~
p^z=m_{\perp}\sinh\theta.
\label{eq:E17}
\end{equation}

At large $m_{\perp}\tau$, the phase of the wave function $\Xi_{\theta,p_\perp}$
is stationary in a very narrow interval  around $~\phi=\theta~$
(outside of this interval, the
function reveals oscillations with exponentially increasing  frequency):
the wave function describes a particle with rapidity $~\theta$. However, for
$m_{\perp}\tau\ll 1$, the phase of the wave function is almost constant along
the surface $\tau=const$~.  The smaller $\tau$, the more uniformly the
particle is spread along the light cone. Up to distortions caused by
the finite
size of the interaction domain, any high-energy collision will produce a
distribution which is uniform in rapidity in the vicinity of the light cone.
The picture looks as if the incoming nuclei carry this distribution {\it ab
initio}. The latter is not surprising as the same arguments can be applied
to the states of particles before a strongly localized interaction.
The distribution $dN\sim const\times d\theta$ corresponds to
$dN\sim const\times dx_F/x_F$
in terms of the Feynman variable $x_F$. Thus, we arrive at a
result which is typical for the Williams-Weiszacker  approach. The full
consideration for the QCD-nucleus has been recently  given by McLerran and
Venugopalan \cite{McLer}.

We assume that deviation from this ideal distribution can be studied
perturbatively with increasing accuracy the more inelastic the
collision is. We expect that
the desired information about this deviation can be obtained  from the deep
inelastic electron-proton scattering data.  We argue that  these data do not
imply exact knowledge of the density matrix of a nucleus. In order to
incorporate information obtained from DIS we begin by computing the DIS
cross-section in terms of the quantum fields kinetic (QFK)\cite{Mak}.

 \renewcommand{\theequation}{3.\arabic{equation}}
\setcounter{equation}{0}
\bigskip
{\bf \Large 3. Deep inelastic scattering on the electron}
\bigskip

The goal of this section is to find those elements of a theory which are common
to two essentially different problems, {\it viz.},
nucleus-nucleus (or proton-proton)
collisions and deep inelastic electron-proton scattering. Our primary demand
is that these elements should appear as a by-product of the two independent
lines of calculations, initiated separately from first principles.

We divide this section into two parts. For the sake of completeness we begin
with a brief definition of the DIS cross-section in terms of the QFK-approach
and define the null-plane variables which will be used for all following
calculations.

Before turning to a detailed derivation of the self-consistent equations in
the next Section, we give an instructive example of the lowest order
calculations. These have no direct physical value, but they allow one to
overcome technical problems and avoid
premature discussion of the highly nontrivial approximations.

\bigskip
\bigskip
{\underline{\it 3.1.~Basic definitions for DIS.}}
\bigskip

As was emphasized in the Introduction, it is important to have similar
definitions of observables for all processes which will participate in
the future information exchange. We may rewrite Eq.(\ref{eq:E1})
for the inclusive amplitude of DIS as
 \begin{equation}
 \langle X|a({\bf k'})Sa^{\dag}({\bf k})|in\rangle~ ~ ~ ,
\label{eq:F1}
\end{equation}
  where $k$ and $k'$ are the laboratory frame
 momenta of the electron before and after the
scattering.  If $q=k-k'$ is the space--like momentum transfer, then the
DIS cross-section is given by
\begin{equation}
 k'_{0} {d\sigma \over d{\bf k'}}= {i\alpha \over(4\pi)^2}
{L_{\mu\nu}(k,k')\over (kP)} {W^{\mu\nu}(q) \over (q^2)^2}~ ~ ~,
\label{eq:F2}
\end{equation}
where $W^{\mu\nu}(q)$ is a standard Bjorken notation for the
correlator of  two electromagnetic currents,
\begin{equation}
 W^{\mu\nu}(q)= {2V_{lab}P^{0} \over 4\pi}[-i \pi^{\mu\nu}_{10}(q)]~ ~ ~.
\label{eq:F3}\end{equation}
We accept without any discussion its standard tensor decomposition,
\begin{equation}
 W^{\mu\nu}(q)= e^{\mu\nu}{\nu W_L \over 2x_{Bj} }+
 \zeta^{\mu\nu}{\nu W_2 \over 2x_{Bj} M^2 }~,
\label{eq:F4} \end{equation}
where $\nu = qP$, $Q^2 = -q^2>0$,  $x_{Bj}=Q^2/2\nu$   and
\begin{equation}
e^{\mu\nu}=-g^{\mu\nu}+{q^\mu q^\nu \over q^2 } ; \;\;\;
\zeta^{\mu\nu}=-g^{\mu\nu}+{P^\mu q^\nu+q^\mu P^\nu \over \nu} -
 q^2 {P^\mu P^\nu \over \nu^2 }~.
\label{eq:F5}
\end{equation}
   Hereafter we will perform all computations using the infinite momentum
frame fixed by the null-plane vector $n^{\mu}$,
\begin{equation}
    n^\mu =(1,{\bf 0_t},-1),\;\;\;\; n^2 =0.
\label{eq:F6}
\end{equation}
It defines the ``+''-components of the Lorentz vectors,
\begin{eqnarray}
    na = a^+ = a_- = a^0 + a^3;\;\;\;\;  a^- = a_+ = a^0 - a^3  \nonumber
\end{eqnarray}
In the infinite momentum frame, the 4-vector of the proton's momentum has
components
\begin{equation}
     P^\mu=(P^+/2,{\bf 0_t},P^+/2),\;\;\;\;\; P^-=P^0-P^3=0.
\label{eq:F7}
\end{equation}
The momentum transfer has the components
\begin{equation}
   q^\mu=(\nu/P^+,{\bf q_t},-\nu/P^+),\;\;\;\; q^+=0, \;\;\;q^-=2\nu/P^+.
\label{eq:F8}
\end{equation}

Instead of the invariant $W_2$,
we will use the mass-independent structure function
 $F_2(x_{Bj},Q^2)= \nu W_2 / M^2$,  which is calculated via the equation
\begin{equation}
c_2=  W^{\mu\nu}n_\mu n_\nu ={(P^+)^2 F_2 \over \nu}.
\label{eq:F9}
\end{equation}

The
longitudinal structure function $F_L(x_{Bj},Q2)=W_L$ should be calculated in
accordance with
\begin{equation}
3F_L = 2x_{Bj}c_1 + 2F_2;\;\;\;\;   c_1 =  W^{\mu\nu}g_{\mu\nu}.
\label{eq:F10}
\end{equation}

\bigskip
\bigskip
{\underline{\it 3.2.~An instructive example: the low order calculation.}}
\bigskip

    In order to proceed with the calculations we must specify the
density matrix.
We shall begin the physical motivation of our choice by reminding the reader
that a widely used approach based on the Wilson's operator product expansion
(OPE) does not utilize any information about the proton's internal structure.
Only the total momentum and the discrete quantum numbers are controlled by
sum rules.  Indeed, the dynamical equations of  QCD contribute only
to the singular coefficient functions while the regular operator functions, the
averages over the proton's state, remain unknown. We can only decide whether
or not to
include the high twist operators in the expansion. The twist-one operators
of OPE correspond to one-particle matrix elements of the proton's density
matrix.  Including the twist-two operators into  OPE would correspond to
irreducible two-particle correlations in the density matrix used here.

The system of integral equations which we expect to derive eventually does not
require any explicit form of the density matrix either. Nevertheless it is
useful to keep in mind some representation
which may serve as a simple reference point.
For example, we may chose an artificial exponential form which reproduces the
total momentum flux of the proton and allows to derive the integral equations
of the Schwinger-Dyson type\cite{Mak}.

    The twist-one operator functions of OPE, by their structure, are binary
products of quark and gluon fields and to some extent resemble occupation
numbers which enter the on-mass-shell Greenians. For example, in the
statistical ensemble we usually have
\begin{equation}
     G_{10}(p)  = -2 \pi i (\not p +m)\delta(p^{2}-m^{2})
     [\theta (p_{0})(1-n^{(+)}(p))- \theta (-p_{0})n^{(-)}(p)].
\label{eq:F11}\end{equation}

    We define matrix elements of our one-particle (twist-one) density matrix
by a certain set of the Greenians. We assign the
superscript ``$\# $'' to all states
in the continuum of free on-mass-shell fields:
\begin{eqnarray}
 G^{\# ij}_{{\stackrel{10}{\scriptscriptstyle 01}}}(p)  =
   -2 \pi i \delta_{ij}(\not p +m)
\theta (\pm p^{0})\delta (p^{2}-m^{2}),  \\ \label{eq:F12}
 D^{\#ab,\mu \nu}_{{\stackrel{10}{\scriptscriptstyle 01}}}(p)  =
     -2 \pi i   \delta_{ab} d^{\mu\nu}(p)
\theta (\pm p_{0})\delta (p^{2}) .
\label{eq:F13}\end{eqnarray}
These states are initially empty and the vacuum correlators $G^{\#}_{10,01}(p)$
 and $D^{\#}_{10,01}(p)$ represent only on-mass-shell particles in
the final states.

  The superscript ``$\ast$'' will label ``bounded'' states of
``valence'' quarks and gluons in the initial proton:
\begin{equation}
{ G^*}_{01}(p)= 2\pi i {(2\pi)^3 \over V_{lab}}
{1\over 3}\delta_{ij} {1\over 2} \not p \delta(p^2)\delta({\bf p}_t)
\theta(p^+) {\cal V}(p^+),
\label{eq:F14}
\end{equation}
\begin{equation}
{ G^*}_{10}(p)=0.
\label{eq:F15}
\end{equation}
Equation (\ref{eq:F14}) describes the phenomenological distribution of the
``valence'' quarks as a function of their light-cone momenta $p^+$,
while Eq.~(\ref{eq:F15}) means that there are no ``valence'' anti-quarks
within the proton. (This assumes that we start at very low scale. Otherwise
Eq.~(\ref{eq:F15}) will have the same form
as Eq.~(\ref{eq:F14}).)  The factors 1/2 and 1/3 correspond to  averaging of
the distribution over the quark spin and color, respectively. The factor
$(2\pi)^3/V_{lab}$ corresponds to  normalization:  we consider a flux
with one proton in a volume $V_{lab}$ per unit time.

    In the same way we define the ``initial'' distribution of  ``valence''
gluons by
\begin{equation}
D^{*ab,\mu \nu}_{{\stackrel{10}{\scriptscriptstyle 01}}}(p)  =
     -2 \pi i{1 \over 8} {(2\pi)^3 \over V_{lab}}
  \delta_{ab} {1 \over 2} d^{\mu\nu}(p) \theta (\pm p_{0})
\delta (p^{2}) \delta({\bf p}_t) \theta(p^+) {\cal G}(p^+),
\label{eq:F16}
\end{equation}
with 1/8 and 1/2 standing
for the color and polarization average and where $d^{\mu\nu}$ is
a projector
\begin{equation}
d^{\mu\nu}(p)=-g^{\mu\nu}+{p^\mu n^\nu+n^\mu p^\nu \over (p^+) }~,
\label{eq:F17}
\end{equation}
which is a sum over the physical gluon polarizations in the null-plane gauge,
\begin{equation}
 n^{\mu}B^{a}_{\mu}= 0;\;\;\;\;   n^2 = 0.
\label{eq:F18}
\end{equation}

The reason for introducing these distributions
is to give definite values to the quantum
numbers (the charges and the momenta). Their densities are given by
\begin{equation}
j^+(p) = Tr \gamma^+ G^{\# ij}_{01}(p) =
{1 \over V_{lab}}\int_{0}^{P^+} d p^+ {\cal V}(p^+)
 ={P^+ \over  V_{lab}} \int_{0}^{1} d x {\cal V}(x)
\label{eq:F19}
\end{equation}
for the quark's light-cone charge flux, and by
\begin{equation}
T^{++}_{q} = -i \int {d^4 p \over (2\pi)^4}
 Tr \gamma^+ p^+ G^{\# ii}_{01}(p) =
{1 \over  V_{lab}} \int_{0}^{P^+} d p^+ p^+ {\cal V}(p^+)
 ={P^+ \over  V_{lab}} \int_{0}^{1} d x x {\cal V}(x)
\label{eq:F20}
\end{equation}
for the (++)-component of the quark energy-momentum tensor. In these
equations
we have introduced the Feynman variable, $x=x_F=p^+/P^+$.
The momentum flux density from the gluon component is given by
\begin{equation}
T_{g}^{++} = -i \int {d^4 p \over (2\pi)^4}
 (p^+)^2 g_{\mu\nu} D^{\ast\mu\nu}_{01}(p)=
{1 \over  V_{lab}} \int_{0}^{P^+} d p^+ p^+ {\cal G}(p^+)
 ={P^+ \over  V_{lab}} \int_{0}^{1} d x x {\cal G}(x)
\label{eq:F21}
\end{equation}
where $P^+{\cal V}(p^+)\equiv {\cal V}(x)$ and
$P^+{\cal G}(p^+)\equiv {\cal G}(x)$.  The initial quark and gluon
distributions are normalized in such a way that in aggregate they carry the
proton's total quantum numbers.

Neglecting any corrections to the electromagnetic vertex we may
rewrite Eq.~(\ref{eq:F2}) in the following way,
\begin{equation}
 W^{\mu\nu}(q)=e_{f}^{2} {2V_{lab}P_{lab} \over 4\pi}
\int {d^4 p \over (2\pi)^4} Tr \gamma^\mu {\bf G}_{10}(p+q) \gamma^\nu
   {\bf G}_{01}(p)~ ~ ~.
\label{eq:F22}
\end{equation}
The off-diagonal quark Greenians (field correlators) in this equation
obey integral  equations
\cite{Mak} which express them in terms of exact retarded and advanced
propagators and sources $\Sigma_{10;01}$ (the ``current'' correlators):
\begin{equation}
{\bf G}_{{\stackrel{10}{\scriptscriptstyle 01}}}
= {\bf G}_{ret}{\stackrel{\leftarrow}{G_{(0)}^{-1}}}
G_{{\stackrel{10}{\scriptscriptstyle 01}}}
{\stackrel{\rightarrow}{G_{(0)}^{-1}}}{\bf G}_{adv}
-{\bf G}_{ret}\Sigma_{{\stackrel{10}{\scriptscriptstyle 01}}}{\bf G}_{adv}.
\label{eq:F23}
\end{equation}

The retarded and advanced Green's functions obey more familiar equations,
\begin{equation}
 {\bf G}_{ret} = G_{ret}+  G_{ret} \Sigma_{ret} {\bf G}_{ret}~ ~ ~,
\label{eq:F24} \end{equation}
\begin{equation}
 {\bf G}_{adv} = G_{adv}+  G_{adv} \Sigma_{adv} {\bf G}_{adv}~ ~ ~,
\label{eq:F25}\end{equation}
which allows a symbolic solution
\begin{equation}
 {\bf G}_{ret}^{-1} = G_{ret}^{-1}-\Sigma_{ret},\;\;\;\;
 {\bf G}_{adv}^{-1} = G_{adv}^{-1}-\Sigma_{adv}~ ~ ~.
\label{eq:F26} \end{equation}
    In the first approximation we may replace the exact retarded and advanced
quark Green's functions, ${\bf G}_{ret}$ and ${\bf G}_{adv}$, by the bare ones
that carry the same leading light-cone singularity,
\begin{equation}
 G_{ret}(p) = {\not p \over (p^0 + i0)^2 - {\bf p}^2 },\;\;\;\;
 G_{adv}(p) = {\not p \over (p^0 - i0)^2 - {\bf p}^2 }.
\label{eq:F27} \end{equation}
Then in virtue of  ~(\ref{eq:F15}), Eqs.~(\ref{eq:F23})
take the following form
 \begin{eqnarray}
{\bf G}_{01}=G^{\#}_{01}+G^{*}_{01}- G_{ret}\Sigma_{01}G_{adv}~ ~ ~, \\
\label{eq:F28}
{\bf G}_{10}=G^{\#}_{10} -  G_{ret}\Sigma_{10} G_{adv}~ ~ ~.
\label{eq:F29}
\end{eqnarray}

In the lowest order we neglect sources   and leave only correlators of
initial fields:
\begin{equation}
 W^{\mu\nu}(q)=e_{f}^{2} {2V_{lab}P_{lab} \over 4\pi}
\int {d^4 p \over (2\pi)^4} Tr \gamma^\mu G^{\#}_{10}(p+q) \gamma^\nu
   G^{*}_{01}(p) ~ ~ ~.
\label{eq:F30}
\end{equation}
Substituting Eqs.~(\ref{eq:F12}) and (\ref{eq:F14}) and rewriting the
delta-function of the on-mass-shell final-state quark as
\begin{equation}
\delta_+((p+q)^2)={1 \over 2\nu} \delta(x-x_{Bj}),
\label{eq:F31}\end{equation}
we get the incredibly simple result of the ``naive'' parton model:
\begin{equation}
F^{(0)}_2(x_{Bj}) =e_{f}^{2} \int_{0}^{1} d x \delta(x-x_{Bj})x {\cal V}(x).
\label{eq:F32}
\end{equation}

In the next approximation we should include the quark fields coming from the
quark and anti-quark sources, $\Sigma_{01}$ and  $\Sigma_{10}$, {\it
i.e.}, the last terms in Eqs.~(\ref{eq:F28}) and
(\ref{eq:F29}). The general expression for the quark
self-energy matrix is given by Eq.~(\ref{eq:E7}). Still restricting ourselves
to bare vertices and bare tree Greenians  we get:
\begin{equation}
 \Sigma^{*\#}_{01}(p)=ig^{2} C_F
\int {d^4 k \over (2\pi)^4}
 \{Tr \gamma_\mu {\bf G}^{*}_{01}(k) \gamma_\nu D^{\#\nu\mu}_{10}(k-p)+
Tr \gamma_\mu G^{\#}_{01}(k) \gamma_\nu{\bf D}^{*\nu\mu}_{10}(k-p),
\label{eq:F33}
\end{equation}
\begin{equation}
 \Sigma^{*\#}_{10}(p)=ig^{2} C_F
\int {d^4 k \over (2\pi)^4}
 \{Tr \gamma_\mu {\bf G}^{\#}_{10}(k+p) \gamma_\nu D^{\ast\nu\mu}_{01}(k).
\label{eq:F34}
\end{equation}

The superscript ``$\ast\#$'' means that one of the contributing states
belongs to the set of
final states, while the other originates from the initial proton.

The gluon correlators obey the following equations:
 \begin{eqnarray}
{\bf D}_{01,10}=D^{\#}_{01,10}+D^{*}_{01,10}- {\bf D}_{ret}\Pi_{01,10}
{\bf D}_{adv} ,
\label{eq:F35}
\end{eqnarray}
\begin{equation}
 {\bf D}_{ret} = D_{ret}+  D_{ret} \Pi_{ret} {\bf D}_{ret},
\label{eq:F36} \end{equation}
\begin{equation}
 {\bf D}_{adv} = D_{adv}+  D_{adv} \Pi_{adv} {\bf D}_{adv}.
\label{eq:F37}\end{equation}
In the first approximation, all contributions of the gluon sources,
$\Pi_{10,01}$,  should be dropped along with radiative corrections to
the retarded and advanced gluon propagators.
The tensor structure of the self-energy may be only of the following form
\begin{equation}
\Sigma(p)=\not p \sigma_2(p)+ \not n \sigma_3(p).
\label{eq:F38}\end{equation}
In all the equations it enters in a single combination,
\begin{eqnarray}
\not p \Sigma(p) \not p=\not p \sigma_1(p) +\not n p^2 p^+ \sigma_3(p);\;\;\;\;
\sigma_1(p)=  p^2 \sigma_2(p)+2 (p^{+})^2 \sigma_3(p).
\label{eq:F39}\end{eqnarray}
After a long, but routine calculation we find that
\begin{eqnarray}
 i\sigma^{*\#}_{1}(p)=-{\pi g^{2} \over V_{lab}P^+ } p^2
\int_{p^+}^{P^+} {d k^+ \over k^+}\delta[(p^+-k^+)p^- -p_{t}^{2}]
{k^+ \over p^+} \{ C_F{z^2+1 \over 1-z}  P^+ {\cal V}(k^+) +
{2 z^2 -2z +1 \over 2} P^+ {\cal G}(k^+) \} ,
\label{eq:F40}\end{eqnarray}
\begin{equation}
 i\sigma^{*\#}_{2}(p)={\pi g^{2} \over V_{lab}P^+ }
\int_{p^+}^{P^+} {d k^+ \over k^+}\delta[(p^+-k^+)p^- -p_{t}^{2}]
{k^+ \over p^+}  \{ C_F P^+ {\cal V}(k^+) +(1-z) P^+ {\cal G}(k^+) \},
\label{eq:F41}\end{equation}
where $z = p^+ /k^+$, and the invariants for the
anti-quark source $\Sigma_{10}$
do not contain terms with valence distributions ${\cal V}(k^+)$.

It is now straightforward  to find the first correction
$F_{2}^{(1)}(x_{Bj},Q^2)$ to the DIS structure function:
\begin{equation}
F^{(1)}_2(x_{Bj}) =e_{f}^{2} \int_{0}^{1} d x \delta(x-x_{Bj})x
 \Delta^{(1)} q_f(x).
\label{eq:F42}\end{equation}
It is presented in the same form as the zero-order term, (\ref{eq:F32}), with
${\cal V}_f(x) =q_f(x, Q^{2}_{0})$ replaced by
\begin{equation}
 \Delta^{(1)} q_f(x,Q^2)={V_{lab}P^+ \over (2\pi)^3}
\int_{Q_{0}^{2}}^{Q^{2}} dp_{t}^{2}
\int dp^{-} p^+ {i\sigma^{*\#}_{1}(p) \over [p^2]^2 }.
\label{eq:F43}\end{equation}
 To be consistent with the resonant condition of the measurement
{}~(\ref{eq:F31}), we must
require that $Q^2$ is large enough (formally, $Q^2\rightarrow \infty$),
and that the behavior of the integrand at high $p_{t}^{2}$ guarantees the
convergence of the integral.
If we substitute ~(\ref{eq:F41}) into the r.h.s.  of ~(\ref{eq:F43})
and perform
a residual integration over $p^-$ using the delta-function:
\begin{equation}
\int dp^{-}\theta(k^+-p^+) {\delta[(p^+-k^+)p^- -p_{t}^{2}] \over p^2 }
=-{1 \over k^+p_{t}^{2}},
\label{eq:F44}\end{equation}
then we easily recover the
first approximation of the Altarelli-Parisi equation for
the non-singlet quark structure functions of the deep inelastic electron-proton
scattering.

    In the next order we must iterate Eq.~(\ref{eq:F35}), including the
influence of the source  $\Pi^{\mu\nu}_{01}$ on the gluon field.
Cutting the accuracy of calculations in Eq.~(\ref{eq:E11}) to bare
vertices, and
neglecting the sources in the internal Greenians, we get the result
\begin{eqnarray}
  \Pi^{\ast\#\mu\nu}_{01}(p)= ig_{r}^{2}
\{\!\int \! {d^4 k \over (2\pi)^4}
 Tr[\gamma^{\mu}{\bf G}^{*}_{01}(k) \gamma^{\nu} G^{\#}_{10}(k-p) +
\gamma^{\mu} G^{\#}_{01}(k+p)\gamma^{\nu}{\bf G}^{*}_{10}(k)]- \nonumber \\
-\int  {d^4 k \over (2\pi)^4}
  V^{\mu\alpha\lambda}_{acf}(p,k-p,-k)
{\bf D}_{01,cc'}^{*\alpha\beta}(k)
 V^{\nu\beta\sigma}_{bc'f'}(-p,p-k,k)
 D_{10, f'f}^{\#\sigma\lambda}(k-p)] \}~~,
\label{eq:F45}\end{eqnarray}
where a sum over the quark flavor $f$ is assumed in the first term.

The polarization tensor $\Pi^{\mu\nu}$  only appears between retarded and
advanced propagators: $[D_{ret}(p)\Pi (p)D_{adv}(p)]^{\mu\nu}$.
The latter contain projectors $d^{\mu\nu}(p)$
which are orthogonal to the 4-vector $n^{\mu}$. So of
the general tensor, only two terms survive:
\begin{equation}
 \Pi^{\mu\nu}(p) =g^{\mu\nu}w_1(p)+{p^\mu p^\nu \over p^2}w_2(p).
\label{eq:F46}\end{equation}
Others, like $p^{\mu}n^{\nu}+ n^{\mu}p^{\nu}$  or $n^{\mu}n^{\nu}$ ,
will cancel out.  Introducing one more projector,
\begin{equation}
{\bar d}^{\mu\nu}(p)=- d^{\mu\rho}(p) d_{\rho}^{\nu}(p)
=-g^{\mu\nu}+{p^\mu n^\nu+n^\mu p^\nu \over (np) }
 -p^2{n^\mu n^\nu \over  (p^+)^2} ,
\label{eq:F47}\end{equation}
which is orthogonal to both vectors $n^{\nu}$ and $p^{\mu}$, we find that
the invariants $w_1$ and $w_2$ can be found from two convolutions,
\begin{equation}
-{\bar d}_{\mu\nu}(p) \Pi^{\mu\nu}(p) =2 w_1(p) ; \;\;\;\;\;
n_\mu n_\nu \Pi^{\mu\nu}(p)={(p^+)^2 \over p^2} w_2(p),
\label{eq:F48}\end{equation}
independently of other invariants accompanying the missing tensor
structures.  The new projector, which includes only two transversal
gluon modes, naturally appears in the tensor with a gluon source~:
\begin{equation}
[d(p)\Pi(p)d(p)]^{\mu\nu}~  = -{\bar d}_{\mu\nu}(p) w_1(p)+
{(p^+)^2 \over p^2} w_2(p)n^\mu n^\nu .
\label{eq:F49}\end{equation}
Now it is easy to find the first approximation for the invariants
$w_1(p)$  and $w_2(p)$~:
\begin{eqnarray}
 iw^{*\#}_{1}(p)=-{\pi g^{2} \over V_{lab}P^+ } p^2
\int_{p^+}^{P^+} {d k^+ \over k^+}\delta[(p^+-k^+)p^- -p_{t}^{2}]
       {k^+ \over p^+} \times\hspace*{1cm}  \nonumber \\
  \times \{ C_F{1+(1-z)^2 \over z}  P^+ {\cal V}(k^+) +
2N_c[z(1-z)+ {z \over 1-z}+{1-z \over z} ] P^+ {\cal G}(k^+) \},
\label{eq:F50}\end{eqnarray}
 \begin{eqnarray}
 iw^{*\#}_{2}(p)=-{\pi g^{2} \over V_{lab}P^+ } p^2
\int_{p^+}^{P^+} {d k^+ \over k^+}\delta[(p^+-k^+)p^- -p_{t}^{2}]
{k^+ \over p^+} \{4 C_F{1-z \over z}  P^+ {\cal V}(k^+)
+ 4 N_c z(1-{z\over 2}) P^+ {\cal G}(k^+) \}.
\label{eq:F51}\end{eqnarray}
In accordance with the previous convention,
and as a reminder of the approximations involved,
the invariants
carry superscripts $\ast\#$~. These indicate that the invariants
are contributed to by one proton's ``bound''
state and one on-mass-shell final state
of a quark or gluon. We hope that the reader is not confused by the absence of
other indices like quark color, or indices indicating the type of ordering
in the invariants $w_i$ and $\sigma_i$.  They can easily be recovered when
it is needed. For example, $w_i$ and $\sigma_i$ are parts of the self-energy
which are summed over the  color. So if $\Pi$   and $\Sigma$  appear as
internal elements in any formula,
we must restore the color factors in the following way:
\begin{eqnarray}
\Sigma\rightarrow\Sigma_{ij}={\delta_{ij} \over 3}\Sigma,\;\;\;\;
\Pi\rightarrow\Pi_{ab}={\delta_{ab} \over 8}\Pi. \nonumber
\end{eqnarray}

Now we may reconstruct a missing element, {\it viz.}, the first
correction to the gluon structure function,
\begin{equation}
 \Delta^{(1)} G(x,Q^2)={V_{lab}P^+ \over (2\pi)^3}
\int_{Q_{0}^{2}}^{Q^{2}} dp_{t}^{2}
\int dp^{-} p^+ {iw^{*\#}_{1}(p) \over [p^2]^2 },
\label{eq:F52}\end{equation}
which is similar to the correction (\ref{eq:F44})
to the quark structure function. In a sequence of approximations it should be
added to the ``valence'' gluon distribution ${\cal G}(x)$.

If we substitute Eq.~(\ref{eq:F50}) into
(\ref{eq:F52}), we immediately obtain
the lowest order approximation of the second Altarelli-Parisi equation for the
 gluon structure  function of the deep inelastic electron-proton scattering.

Concluding this section, let us pay special attention to the infrared
poles at $z = 1$,
which originate from the pinch-poles of the gluon Greenians in
the null-plane gauge. To cure this problem we will proceed following Altarelli
and Parisi \cite{AP}.
We will first shield the IR singularity by introduction
the ``plus-distributions,''
\begin{eqnarray}
\int_{0}^{1}{f(z)dz\over (1-z)_+}=\int_{0}^{1}{f(z)-f(1)\over 1-z}dz,\nonumber
\end{eqnarray}
and then modify the end-point behavior in such a way that the first integrals
could not be changed by the radiative corrections.
The first integrals are the total flux of the flavor $f$,
\begin{equation}
j^{+}_f = -iV_{lab}\int {d^4p\over (2\pi)^4} {\rm Tr} \gamma^+
[{\bf G}^{f}_{01}(p)-{\bf G}^{f}_{10}(p)]~ ~ ~,
\label{eq:F53}\end{equation}
and the total flux of the light-cone momentum,
\begin{equation}
T^{++} = -iV_{lab}\int {d^4 p \over (2\pi)^4} \{\sum_{f}
 {\rm Tr} \gamma^+ p^+ [{\bf G}^{f}_{01}(p)+{\bf G}^{f}_{10}(p)]+
(p^+)^2g_{\mu\nu} {\bf D}^{\mu\nu}_{01}(p) \}~ ~ ~.
\label{eq:F54}\end{equation}
As follows from Eqs.~(\ref{eq:F43}) and (\ref{eq:F52}), the first
radiative corrections to the flavor and momentum fluxes are
\begin{equation}
\Delta j^{+}_f = {P^+V_{lab}\over (2\pi)^3} \int_{0}^{P^+} dp^+
\int_{Q_{0}^{2}}^{Q^{2}}dp_{t}^{2}\int dp^- p^+
\left[{i\sigma^{f,01}_{1}(p) \over  [p^2]^2}-
{i\sigma^{f,10}_{1}(p) \over  [p^2]^2} \right]
\label{eq:F55}\end{equation}
and
\begin{eqnarray}
\Delta T^{++} \!= {P^+V_{lab}\over (2\pi)^3} \int_{0}^{P^+}\!\! dp^+ \!\!
\int_{Q_{0}^{2}}^{Q^{2}}\!\! dp_{t}^{2}\int\! dp^- (p^+)^2 \left[\sum_{f}
[{i\sigma^{f,01}_{1}(p) \over  [p^2]^2}+
{i\sigma^{f,10}_{1}(p) \over  [p^2]^2} ]+
{iw^{01}_{1}(p) \over  [p^2]^2}   \right],
\label{eq:F56}
\end{eqnarray}
respectively. The superscript $``\#\ast''$ is omitted because these equations
remain valid beyond
the first order approximation. The resulting conditions, $\Delta j^{+}_f=0$
and $\Delta T^{++}=0$,  for the splitting kernels
in the leading logarithmic approximation are obvious,
\begin{eqnarray}
 \int_{0}^{1} P_{qq}(z)dz = 0,   \nonumber\\
 \int_{0}^{1}[zP_{gg}(z) + 2n_f z P_{qg}(z)]dz = 0  \\ \label{eq:F57}
 \int_{0}^{1}[zP_{gq}(z) + zP_{qq}(z)]dz = 0.  \nonumber
\end{eqnarray}
However, beyond the LLA they may change.
One now readily finds an explicit form of the splitting kernels,
\begin{eqnarray}
P_{qq}(z)=C_F\left[{z^2+1\over (1-z)_+}
+{3\over 2}\delta (1-z)\right], \nonumber\\
P_{qg}(z)= {z^2+(1-z)^2 \over 2}, \;\;\;
P_{gq}(z) = C_F {1+(1-z^2) \over z}  \\ \label{eq:F58}
P_{gg}(z) = 2N_c\left[z(1-z)+ {z \over (1-z)_+}+{1-z \over z}
+{\beta_0 \over 4N_c}\delta (1-z) \right] ~ ~ ~,    \nonumber
\end{eqnarray}
where the factor $\beta_{0} = 11 - 2n_f/3$ coincides with the first
coefficient of the Gell-Mann-Low function.

 \renewcommand{\theequation}{4.\arabic{equation}}
\setcounter{equation}{0}
\bigskip
{\bf \Large 4. Evolution of the sources.}
\bigskip

Now we are ready to derive integral equations that govern the field
correlators and their sources.  Actually, they have already been given above.
An examination
of the calculations in the previous Section shows that we did not sum any
series. We were consequently performing a series expansion of
the previously derived
self-consistent solution of the integral Schwinger-Dyson equations.
These are Eqs.~(\ref{eq:E8}) and (\ref{eq:E11}), which define the
self-energies via the Greenians, and Eqs.~(\ref{eq:F28}), (\ref{eq:F29})
and  (\ref{eq:F35}), which define the Greenians (field correlators)
via the self-energies. We shall rewrite the last equations in the form
\begin{eqnarray}
{\bf G}_{10,01}= G^{\#}_{10,01}-{\bf G}_{ret}\Sigma_{10,01}{\bf
G}_{adv}~ ~ ~,~ ~{\rm and}  \\  \label{eq:G1}
{\bf D}_{10,01}= D^{\#}_{10,01}-{\bf D}_{ret}\Pi_{10,01}{\bf D}_{adv}~~,
\label{eq:G2}\end{eqnarray}
omitting the $\ast$-labelled terms as they do not contribute to
the differential
form  of evolution equations. In contrast to the integral evolution equations,
the differential equations do not require any information about the initial
data. At this point we do not make any approximations.

\bigskip
\bigskip
{\underline{\it 4.1.~Dynamical equations in the leading
logarithmic approximation.}}
\bigskip

In order to obtain the equations of the leading logarithmic
approximation, which sum up the perturbation series with the
leading logarithms, we must consider the vertex operators in
Eqs.~(\ref{eq:E8}) and (\ref{eq:E11}) as the bare ones. We must also
confine one of the off-diagonal field correlators to the out-states in
the continuum~:
\begin{eqnarray}
 \Sigma_{01}(p)=ig^{2} C_F \!\!
\int \!\! {d^4 k \over (2\pi)^4}
 {\rm Tr} \{ \gamma_\mu G_{ret}(k) \Sigma_{01}(k) G_{adv}(k)
 \gamma_\nu D^{\#\nu\mu}_{10}(k-p)
+ \gamma_\mu G^{\#}_{01}(k+p) \gamma_\nu
\left[D_{ret}(k) \Pi_{10}(k) D_{adv}(k)\right]^{\nu\mu}\}
\label{eq:G3}\end{eqnarray}
\begin{eqnarray}
  \Pi^{\mu\nu}_{01}(p)=-ig_{r}^{2}
\{-\int \! {d^4 k \over (2\pi)^4}
{\rm Tr}[\gamma^{\mu}G_{ret}(k) \Sigma_{01}(k) G_{adv}(k)
 \gamma^{\nu} G^{\#}_{10}(k-p)
+\gamma^{\mu} G^{\#}_{01}(k+p)\gamma^{\nu}
G_{ret}(k) \Sigma_{01}(k) G_{adv}(k)]+  \nonumber \\
+\int  {d^4 k \over (2\pi)^4}
  V^{\mu\alpha\nu}_{acf}(p,k-p,k)
\left[D_{ret}(k) \Pi_{01}(k) D_{adv}(k)\right]^{\alpha\beta}_{cc'}
 V^{\nu\beta\sigma}_{bc'f'}(-p,p-k,-k)
 D_{10, f'f}^{\#\lambda\sigma}(k-p) \}
\label{eq:G4}\end{eqnarray}

By inspection, these equations reveal an astonishing result -
the equations which govern the dynamics of the sources
$\Sigma_{01}$ and  $\Pi_{01}$  of the field correlators ${\bf G}_{01}$
and ${\bf D}_{01}$ have a ladder structure.  This result appeared
though we did
not try to set any momentum or angular ordering of the emission processes.

This result deserves special discussion.  First, let us trace through
the previous calculations once more.  We started from the basic
definitions (\ref{eq:F1}) and
(\ref{eq:F2}) of the observable cross-section of the
DIS. Then we expressed a specific off-diagonal polarization tensor of a proton
via the off-diagonal Greenians ${\bf G}_{10,01}$ and  ${\bf D}_{10,01}$
of the ``kinetic'' technique. Afterwards we
used the Schwinger-Dyson equations for these Greenians to express them via
their sources $\Sigma_{10,01}$ and $\Pi_{10,01}$. This resulted in a closed
system of ladder-type equations for these sources.

Now we must answer two questions: (i) what was the physical input; and (ii)
what follows from the application of new method to the well-known
problem of DIS, for which the solution is already known.

As in any quantum-mechanic problem, the physical picture is specified by a
density matrix of the initial state. Our density matrix carries the same
information as the operator functions of the OPE, that is, no detailed
information. Only the global quantum numbers are under its control.

A more careful analysis
shows that our result is not exactly the same as that of the
OPE-based approach. Though we get (up to a quantitatively inessential shift of
the singularities) the leading logarithm approximation resulting in
the Gribov-Lipatov-Altarelli-Parisi
equations, which sums the leading logarithms of the
perturbation series, an important qualitative difference appears.
The Feynman tree propagators were replaced by the retarded ones.
This change reveals the causal structure of the whole
process.  The last interaction, which
puts a single quark onto its mass shell in the perturbative
vacuum, is also the latest in time. In other
words, the last interaction results
in a collapse of the initial wave function.

So, as a by-product, we have answered an old question
\cite{Coll,Durand,Muller2} about the correspondence
between the evolutionary scale $Q^2$ or $x$ and the temporal scale. The
causal space-time structure is contained, {\it a priori}, in  evolution
equations like the GLAP equations. It is not necessary to impose it
{\it a posteriori}.
The $x$--ordering is a further
consequence  of the $\theta$--functions that allow only for the emission into
the initially unpopulated continuum. The $Q^2$-ordering with respect to
transverse momentum is not a necessary condition, and we shall discuss it
shortly.

What are the practical consequences of this picture? Firstly, in order
to find the structure functions of the deep inelastic scattering of a
proton off an electron,
one should know the intensity of the quark field source
$\Sigma_{01}$ {\it before} the field interaction with the electron.
That is why it does not matter before  {\it what kind}  of interaction.
The intensities of the sources $\Sigma$  and $\Pi$ turn out to be  universal
functions. Unlike the structure functions of DIS they do not depend from
the particular choice of the measurement procedure.
We can continue this reasoning in application to pp- and
AA-collisions. This leads to the conclusion that if we wish to use
e-p DIS data for the description of p-p collision dynamics, we should
rely on the sources $\Sigma$  and $\Pi$, rather than
the structure functions $q(x,Q^2)$ and $G(x,Q^2)$. Their evolution is only a
specific projection of the more complicated dynamic of the sources.

The second consequence is that after the structure functions $q(x,Q^2)$ and
$G(x,Q^2)$ of the e-p DIS are found (simply by fitting data, for example),
we do not need their phenomenological interpretation as a parton density in
order to apply them to other types of collision.

It is well known that instead of calculating the observable
e-p cross-section, the OPE method
computes the imaginary part of the truncated Feynman
amplitude of the auxiliary Compton process. The next step is a renormalization
group analysis of this definite S-matrix amplitude. Unfortunately, the power of
this method is restricted to one single problem. Any extension of this method
requires a parton language.

Indeed, though the cross-section of the Drell-Yan process
is defined by the same (except for kinematic region) polarization
operator as in DIS, it can not be calculated via OPE. The
difference is that now the
operator functions should be averaged over a state with two protons.
Feynman  propagators which contribute to an auxiliary S-matrix amplitude
do not disappear outside the light cone. For the massless partons, and
especially in the infinite momentum frame, it leads to the effective
interaction {\it before} the collision. The factorization theorem \cite{Fact}
may be a remedy, but it requires parton language.
At the same time it is clear that protons colliding at high energies
are causally independent until the moment of collision.

We shall now show that the above ladder equations (\ref{eq:G3})
and (\ref{eq:G4}) are equivalent to the well known QCD evolution equations.
Indeed, let us rewrite Eqs.~(\ref{eq:G3})
and (\ref{eq:G4}) in terms of their tensor components.  Beyond
the first order calculations of Section~2.3. we must also take into account
radiative corrections to the retarded and advanced Green's functions.
The spinor (tensor) structure of $\Sigma_{ret,adv}$ ($\Pi_{ret,adv}$), as
it is given by Eqs.~(\ref{eq:F38}) and (\ref{eq:F46}), remains
unchanged.  The solution of the Schwinger-Dyson equations (\ref{eq:F24}),
(\ref{eq:F25}) and ~(\ref{eq:F36}), (\ref{eq:F37}) for  retarded and
advanced Green's  functions is easily cast in the form:
\begin{eqnarray}
{\bf G}_{ret,adv}(p)&=&{\not p \over  p^2 - \sigma_{1}^{R,A}(p)}-
{\not n p^+ \sigma_{3}^{R,A}(p)\over
(p^2 - \sigma_{1}^{R,A}(p))(1-\sigma_{2}^{R,A}(p))},\nonumber  \\
{\bf D}^{\mu\nu}_{ret,adv}(p)&=&
{{\bar d}^{\mu\nu}(p)\over p^2 - w_{1}^{R,A}(p)}+
{p^2\over (p^+)^2} {n^{\mu}n^{\nu} \over  p^2 - w_{2}^{R,A}(p)}.
\label{eq:G5}\end{eqnarray}

Introducing the following shorthand notations for the
denominators of the propagators of different modes,
\begin{eqnarray}
 {\cal W}^{R,A}_{1}(p) &=& p^2 - w_{1}^{R,A}(p),\;\;\;\;\;\;\;
 {\cal W}^{R,A}_{2}(p) = p^2 - w_{2}^{R,A}(p),\nonumber\\
 {\cal S}^{R,A}_{1}(p)&=& p^2 - \sigma_{1}^{R,A}(p),\;\;\;\;\;\;\;
 {\cal S}^{R,A}_{2}(p) = 1 - \sigma_{2}^{R,A}(p),
\label{eq:G6}\end{eqnarray}
we easily obtain
\begin{eqnarray}
G_{ret}(p) \Sigma_{01}(p) G_{adv}(p)={[\not p -\not
n(p^2/2p^+)]\sigma_{1}^{01}(p)
\over {\cal S}^{R}_{1}(p){\cal S}^{A}_{1}(p)} +{\not p \over 2p^+}
{\sigma_{2}^{01}(p)\over {\cal S}^{R}_{2}(p){\cal S}^{A}_{2}(p)},
\label{eq:G7}\end{eqnarray}
\begin{eqnarray}
\left[D_{ret}(k) \Pi_{01}(k) D_{adv}(k)\right]^{\mu\nu}=
{-{\bar d}^{\mu\nu}(p) w_{1}^{01}(p)\over
{\cal W}^{R}_{1}(p){\cal W}^{A}_{1}(p)}+
{(p^2/(p^+)^2) w_{2}^{01}(p)n^{\mu}n^{\nu}
\over {\cal W}^{R}_{2}(p){\cal W}^{A}_{2}(p)}.
\label{eq:G8}\end{eqnarray}

The complete evolution
equations are long, and are given in Appendix~1 along
with analysis of further approximations. Here we write only the leading
logarithmic terms which eventually result in the GLAP equations:
\begin{eqnarray}
 \sigma_{1}^{01}(p) = {g_{r}^{2} \over (2\pi)^3 }
\int_{p^+}^{P^+} {d k^+\over k^+}\int d^2{\bf k}_{t} d k^-
\delta[(p^+-k^+)(p^- -k^-)-({\bf p}_{t}-{\bf k}_{t})^{2}]\times\nonumber\\
   \left[-p^2{k^+ \over p^+}\right]
\left[ P_{qq}({p^+ \over k^+}){k^+ \sigma_{1}^{01}(k)
 \over {\cal S}^{R}_{1}(k){\cal S}^{A}_{1}(k)  }
 + P_{qg}({p^+ \over k^+}) {k^+ w_{1}^{01}(k) \over
 {\cal W}^{R}_{1}(k){\cal W}^{A}_{1}(k) }+...\right],\hspace*{2cm}
\label{eq:G9}
\end{eqnarray}
\begin{eqnarray}
 w_{1}^{01}(p) = {g_{r}^{2} \over (2\pi)^3}
\int_{p^+}^{P^+} {d k^+\over k^+}\int d^2{\bf k}_{t} d k^-
\delta[(p^+-k^+)(p^- -k^-)-({\bf p}_{t}-{\bf k}_{t})^{2}]\times\nonumber\\
 \times  \left[-p^2{k^+ \over p^+}\right]
\left[ P_{gq}({p^+ \over k^+}){k^+ \sigma_{1}^{01}(k)
 \over {\cal S}^{R}_{1}(k){\cal S}^{A}_{1}(k)  }
 + P_{gg}({p^+ \over k^+}) {k^+ w_{1}^{01}(k) \over
 {\cal W}^{R}_{1}(k){\cal W}^{A}_{1}(k) }+...\right].\hspace*{2cm}
\label{eq:G10}
\end{eqnarray}

    Substituting Eqs.~(\ref{eq:G9}) and (\ref{eq:G10}) into
Eqns.~(\ref{eq:F9}) and (\ref{eq:F22}), we easily find that the well
known structure functions of the electron-proton deep inelastic scattering
are defined by the equations which are similar to
{}~(\ref{eq:F43}) and ~(\ref{eq:F52}):
\begin{equation}
 q_f(x,Q^2)= q_f(x,Q_{0}^{2})+{V_{lab}P^+ \over (2\pi)^3}
\int_{Q_{0}^{2}}^{Q^{2}} dp_{t}^{2}
\int dp^{-}  {ip^+\sigma^{01}_{1}(p) \over
{\cal S}^{R}_{1}(p){\cal S}^{A}_{1}(p) } ,
\label{eq:G11}\end{equation}
\begin{equation}
 G(x,Q^2)= G(x,Q_{0}^{2})+{V_{lab}P^+ \over (2\pi)^3}
\int_{Q_{0}^{2}}^{Q^{2}} dp_{t}^{2} \int dp^{-}
{ip^+ w^{01}_{1}(p)\over {\cal W}^{R}_{1}(p){\cal W}^{A}_{1}(p)}.
\label{eq:G12}\end{equation}
The first of these equations is exact, it does not depend on any further
approximations. The second one is approximate. It holds only in the LLA.

We observe that the structure functions of deep inelastic e-p
scattering in the LLA require only one of the two invariants from each of
the sources $\Pi$ and $\Sigma$. This nice feature fails beyond LLA,
and does not hold for processes
with different polarization properties of the interaction vertex.

    Unfortunately, our knowledge of the confinement dynamics is not sufficient
to proceed without additional qualitative physical input.  We must integrate
over the null-plane momentum $p^-$ and this is a highly nontrivial procedure,
unless we postulate an {\it ad hoc} requirement that the answer should
contain only logarithms.  When we were doing similar calculations in the
lowest order, we  designed the density matrix of ``valence'' quarks and gluons
in such a way that $k^-=0$ and $k_t = 0$.  In doing this, we kept in mind
both the physics of the infinite momentum frame and the proton's confinement
before the collision.

The point is that propagation of quark and gluon fields before they reach a
collision vertex is not free.  Assuming the opposite, we would immediately
violate the causality principle, or be in contradiction with previous
calculations. We shall see shortly that the requirement  of selfconsistency
between emission-absorbtion and propagation provides a natural condition for
renormalization. Suppression of the spectral patterns with $k^-\neq 0$ is  a
part of this condition.
Physically, it means that the field correlators
like $\Pi(x-y)$ do not depend on the difference $x^+-y^+$ of the coordinates in
the direction of the light-cone propagation.
We will conjecture that the exact
retarded and advanced propagators indeed prevent the wave packet of a proton
(or a nucleus) from a premature decay,  and shall cast the requirement in two
forms.

The first, weak form, of this condition  can be cast in the form of the
inequality: $ ~|k^-| << |p^-|$. Then we may integrate over $p^-$ using
the previous formula  (\ref{eq:F44}) as a physical approximation.
The second, strong form replaces the inequality by the exact condition
$p^- = 0$, which can be incorporated into the prescription:
\begin{equation}
{ip^+\sigma^{01}_{1}(p) \over {\cal S}^{R}_{1}(p){\cal S}^{A}_{1}(p) } =
{ (2\pi)^3\over V_{lab}P^+ }\delta(p^{-}){dq_f(x,p_{t}^{2}) \over d p_{t}^{2}},
\label{eq:G14}\end{equation}
\begin{equation}
 {ip^+ w^{01}_{1}(p)\over {\cal W}^{R}_{1}(p){\cal W}^{A}_{1}(p)}=
{ (2\pi)^3\over V_{lab}P^+ }\delta(p^{-}){dG(x,p_{t}^{2}) \over d p_{t}^{2}}.
\label{eq:G15}\end{equation}

These equations are a recipe on how to use DIS data in the
leading logarithmic
approximation.  They require an additional comment about the
 $p_t^2$-integration, which is the next step in obtaining the
DIS structure functions. As we have already argued, an unambiguous definition
of the structure functions is possible only in the limit of
$Q^2,\nu \rightarrow\infty$ which eventually leads to the resonant
condition ~(\ref{eq:F42}) of the measurement : $~x_F=x_{Bj}$.

When the r.h.s. of the Eqs.~(\ref{eq:G9}) and Eq.~(\ref{eq:G10})  are
integrated over $p_{t}^{2}$ with a large upper limit $Q^2$, the
kernel of the resulting equation depends on $k_{t}^{2}$ only in the combination
$k_{t}^{2}+(k^+/p^+)Q^2$.  The $k_{t}^{2}$-behavior of the structure
function in the new integral should guarantee its convergence. Therefore,
one may  neglect  $k_{t}^{2}$ from the very beginning by
assuming that only the domain $k_t\ll p_t$ contributes in the
the initial equations.  This
condition is known as  ``ordering by angles.'' Unlike the ordering by
Feynman $x$, it is not a fundamental requirement. Clearly, this approximation
is not valid at very low $x$. Nevertheless, even for low $x$ the initial
equations ~(\ref{eq:G9}) and ~(\ref{eq:G10}) remain unchanged. In the
Appendix~1, we show that at low $x$ they may be reduced to the BFKL equation.

An explicit expression for the longitudinal structure function of the DIS
follows from Eq.~(\ref{eq:F10}):
\begin{equation}
 3Q^2F_L(x,Q^2)={V_{lab}P^+ \over (2\pi)^3} \sum_{f}e_{f}^{2}
 \int_{Q_{0}^{2}}^{Q^{2}} dp_{t}^{2} \int dp^{-}
\left[{p^2\sigma_{1}(p)\over {\cal S}^{R}_{1}(p){\cal S}^{A}_{1}(p)}
- {\sigma_{2}(p)\over {\cal S}^{R}_{2}(p){\cal S}^{A}_{2}(p)}\right].
\label{eq:G16}\end{equation}
This equation is exact and it proves the Callan-Gross relation when we neglect
the scaling violation ($Q^2$-dependence) in the structure functions.

Up until now, we can only trace the correspondence between our approximate
equations  (\ref{eq:G9}) and  (\ref{eq:G10}), and the LLA of the
OPE-based calculations or the Lipatov's LL(1/x) approximation of the Regge
calculus.  The longitudinal structure function found via OPE has
an extra small factor $\alpha_s$ which is not compensated for by any big
logarithm.  Correspondence between the two approaches in the next orders is
still unclear.  In what follows we are going to use standard set of structure
functions derived from the data in the leading logarithm approximations.
Thus, we shall neglect all terms which can be explicitly reduced to $F_L$.

\bigskip
\bigskip
{\underline{\it 4.2.~Renormalization of the evolution equations. }}
\bigskip

    Until now we have dealt only with objects which do not require
renormalization. All these objects were tightly connected with observables.
Corresponding Greenians and self-energies were imaginary and for this
reason could not contain ultra-violet divergencies. So we could safely
use a ``naive'' form of Schwinger-Dyson equations,
\begin{eqnarray}
  {\bf G}_{AB} = G_{AB}+ \sum_{RS} G_{AR} \Sigma_{RS} {\bf G}_{SB},\;\;\;\;\;
  {\bf D}_{AB} = D_{AB}+ \sum_{RS} D_{AR} \Pi_{RS}{\bf D}_{SB}.
\label{eq:G17}
\end{eqnarray}
The divergent retarded and advanced self-energies were completely neglected,
and retarded and advanced Green's functions were considered as the bare ones.
Consequently, the counter terms of the
Lagrangian still did not manifest themselves, and the  renormalized coupling
constant $g_r$ still remains undefined. We must now fill this gap.

After including the counter-terms we obtain the same equations, but with the
self-energies modified by the quasi-local terms:
\begin{eqnarray}
  {\bf D}_{AB} = D_{AB}+ \sum_{RS} D_{AR}[(Z_{1F} \Pi'_{RS}+Z_1 \Pi''_{RS})+
(1-Z_3)(-1)^R \delta_{RS}D_{0}^{-1}   ] {\bf D}_{SB}~,\\ \label{eq:G19}
  {\bf G}_{AB} = G_{AB}+ \sum_{RS} G_{AR}[Z_{1F}\Sigma_{RS}
+ (1-Z_2)(-1)^R \delta_{RS}G_{0}^{-1}  ] {\bf G}_{SB}~ ~ ~,\label{eq:G20}
\end{eqnarray}
where $ \Pi'$ and $ \Pi''$ are the fermion and gluon loops respectively.
In perturbative calculations the factors $Z_{1,1F}$  should
be split further as $~Z_{1,1F}=1+(Z_{1,1F}-1)~$, with the second term
assigned to the UV-renormalization of the vertex.
The only changes from Eqs.~(\ref{eq:G17}) are
due to additional {\it diagonal} terms in $\Pi$ and $\Sigma$.
This is quite natural as the off-diagonal terms are imaginary and if they were
divergent we would have no remedy to cure the problem.
As the matrix structure of Eqs.~(\ref{eq:G19}) and
(\ref{eq:G20}) is literally the same as that of
Eqs.~(\ref{eq:G17}),  we can rotate the $2\times 2$ basis
as usual \cite{Keld,Mak}.
Keeping in mind the light-cone dominance, we may rewrite equations for
$T$-ordered and $T^{\dag}$-ordered correlators as
\begin{equation}
 {\bf D}_{{\stackrel{00}{\scriptscriptstyle 11}}}(p)
= D^{\#}_{{\stackrel{00}{\scriptscriptstyle 11}}}(p)
  +D_{ret}(p) [(Z_{1F} \Pi'_{{\stackrel{11}{\scriptscriptstyle 00}}}+
Z_1 \Pi''_{{\stackrel{11}{\scriptscriptstyle 00}}})\pm
(1-Z_3)D_{0}^{-1}(p)] D_{adv}(p)~,
 \label{eq:G21}
\end{equation}
\begin{equation}
 {\bf G}_{{\stackrel{00}{\scriptscriptstyle 11}}}(p)
= G^{\#}_{{\stackrel{00}{\scriptscriptstyle 11}}}(p)
  +G_{ret}(p) [Z_{1F}\Sigma_{{\stackrel{11}{\scriptscriptstyle 00}}}(p)
\pm (1-Z_2)G_{0}^{-1}(p)] G_{adv}(p)~,
 \label{eq:G22}
\end{equation}
where $\Pi_{11,00}(p)$ and $\Sigma_{11,00}(p)$ should be calculated using
Eqs.~(\ref{eq:E8}) and ~(\ref{eq:E11}) with the bare vertices and no more
than one source.   We omit the $\ast$-labelled terms here as we are interested
in the UV-renormalization and low-$x$ effects.
These terms are effectively cut off at very high momenta and do not affect
the latest stages of evolution.

Eqs.~(\ref{eq:G21}) -- (\ref{eq:G22})
are too approximate to give an explicit value of running
coupling constant. The questions we want to consider are: (i) does
renormalization of the sources $\Pi$ and $\Sigma$ result in renormalization
of the coupling constant? and (ii) does the coupling constant attached to the
vertex at some moment $t$ require for its renormalization any information
except for the dynamics of the ladder at previous moments?

    Before turning to explicit calculations, let us address some qualitative
issues. The new equations are not completely independent from those we have
already studied. Indeed, the sums and the differences
 \begin{eqnarray}
D_1=D_{00}+D_{11}=D_{10}+D_{01},\;\;\; \;
 \Pi_1= \Pi _{00}+ \Pi_{11}=- \Pi_{10}- \Pi_{01}, \nonumber   \\
 D_0=D_{ret}-D_{adv}=D_{10}-D_{01},\;\;\; \;
 \Pi_0= \Pi _{ret}- \Pi_{adv}=- \Pi_{10}+ \Pi_{01}
\label{eq:G25}
\end{eqnarray}
define imaginary parts of $T$- and $T^{\dag}$-ordered, as well as retarded and
advanced, correlators in an interdependent way.
The real parts are  not independent either. Indeed,
 \begin{eqnarray}
2D_s=D_{00}-D_{11}=D_{ret}+D_{adv}=2{\rm Re}D_{00}=-2{\rm Re}D_{11}
=2{\rm Re}D_{ret}=2{\rm Re}D_{adv},  \nonumber \\
2\Pi_s= \Pi _{00}- \Pi_{11}= \Pi_{ret}+ \Pi_{adv}=
2{\rm Re} \Pi _{00}= -2{\rm Re} \Pi _{11}= 2{\rm Re} \Pi _{ret}
=2{\rm Re} \Pi _{adv}.\hspace*{.5cm}
\label{eq:G26}
\end{eqnarray}
If we recall, in addition, that the causality principle connects real and
imaginary parts of the
retarded and advanced correlators by means of dispersion
relations, then we realise that we have practically no choice in the
regularization of the divergent real functions -- we must follow the BHPZ
scheme \cite{Bogol}.

Let us begin by writing down the integral equation for the gluon polarization
correlators in the leading approximation:
\begin{eqnarray}
  \Pi^{\mu\nu}_{00}(p)=\Pi^{(0)\mu\nu}_{00}(p)
-ig_{r}^{2}\int {d^4 k \over (2\pi)^4}Tr[\gamma^{\mu}G_{00}(k-p)\gamma^{\nu}
G_{ret}(k)[Z_{1F}\Sigma_{11}(p)+(1-Z_2)G_{0}^{-1}(k) ]
G_{adv}(k)-  \nonumber   \\
-ig_{r}^{2} \int  {d^4 k \over (2\pi)^4}
  V^{\mu\alpha\nu}_{acf}(p,k-p,-k)
D_{00,cc'}^{\alpha\beta}(k-p)
 V^{\nu\beta\sigma}_{bc'f'}(-p,p-k,k) \nonumber \\
\times\{ D_{ret}(k)[(Z_{1F} \Pi'_{11}+Z_1 \Pi''_{11})(k)+(1-Z_3)D_{0}^{-1}(k)]
D_{adv}(k)\}_{f'f}^{\lambda\sigma}]
\label{eq:G27}
\end{eqnarray}
The corresponding
equation for $\Pi^{\mu\nu}_{11}(p)$ is the complex anti-conjugate of
Eq.~(\ref{eq:G27}),
and can be obtained via replacement the $T$-ordered functions by the
minus $T^{\dag}$-ordered ones
and {\it vice versa}.  Similar integral equations may
be written for the fermion sources.

In  complete agreement with (\ref{eq:G25}) and (\ref{eq:G26})
these equations have ladder structure with retarded behavior.
To lowest order, gluon and fermion Green functions are given by
 \begin{eqnarray}
D^{\mu\nu}_{{\stackrel{00}{\scriptscriptstyle 11}}}(p)
={\pm d^{\mu\nu}(p) \over p^2 \pm i0},\;\;\;\;
G_{{\stackrel{00}{\scriptscriptstyle 11}}}(p) ={\pm \not p \over p^2 \pm i0},
 \label{eq:G28}
\end{eqnarray}
 and $\Pi^{(0)\mu\nu}_{00}(p)$  is the usual ultraviolet-divergent
vacuum gluon polarization  tensor.

For the sake of simplicity, let us consider only the
gluon sector in the leading
approximation. Projecting
Eq.~(\ref{eq:G28}) onto the normal modes, one obtains
\begin{eqnarray}
  w_{1}^{00}(p)= w_{(0)1}^{00}(p)
-{iZ_1 g_{r}^{2} \over 2(2\pi)^4}  \int  {d^4 k  \over (k-p)^2-i0 }
{\cal I}({p^+\over k^+})
{ w_{1}^{11}(k)+(1-Z_3)k^2 \over [k^2]^2}
\label{eq:G29}
\end{eqnarray}
where we have denoted
\begin{eqnarray}
{\cal I}(z)=8[(-p^2/z + k^2)P_{gg}(z) -(k-p)^2(z+1/z-1/4)]~, \nonumber
\end{eqnarray}
and the splitting kernel $P_{gg}$ is the same as in the evolution equation
for a gluon  source $\Pi_{01}$. It is IR-regularized in the same way,
otherwise we would obtain a contradiction with equations (\ref{eq:G25}) and
(\ref{eq:G26}). Alternating $T$-ordered and $T^{\dag}$-ordered
correlators in the ladder rungs is crucial for the subsequent conclusions.

The imaginary part of Eq.~(\ref{eq:G29}) is finite. It can be obtained as the
sum of the two ladder equations for $w^{01}_{1}$ and $w^{10}_{1}$.  The
divergent
real part of the equation in the leading approximation is an equation
for ${\rm Re}w^{00}_{1}={\rm Re} w^{R}_{1}$, {\it i.e.} the real part of the
retarded self-energy. To facilitate the physical
analysis, we shall derive this
equation in another way,
starting with the explicit expression for the retarded
self-energy. Utilizing the identities $\Pi_{ret}=
\Pi_{00}+\Pi_{01}=-\Pi_{10}-\Pi_{11}$ and $\Pi_{adv}=
\Pi_{00}+\Pi_{10}=-\Pi_{01}-\Pi_{11}$, we easily obtain:
\begin{eqnarray}
\Pi^{\mu\nu}_{{\stackrel{ret}{\scriptscriptstyle adv}}}(p)=
{i\over 2}g_{r}^{2}
\int  {d^4 k \over (2\pi)^4} [ V^{\mu\alpha\lambda}_{acf}(p,-k-p,k)
D_{{\stackrel{ret}{\scriptscriptstyle adv}}}^{cc',\alpha\beta}(k+p)
V^{\nu\beta\sigma}_{bc'f'}(-p,p+k,-k)D_{1}^{f'f,\sigma\lambda}(k)+
\nonumber \\
+ V^{\mu\alpha\lambda}_{acf}(p,-k-p,k)
D_{1}^{cc',\alpha\beta}(k+p)V^{\nu\beta\sigma}_{bc'f'}(-p,p+k,-k)
D_{{\stackrel{adv}{\scriptscriptstyle ret}}}^{f'f,\sigma\lambda}(k)]
\hspace*{1cm}
\label{eq:G23A}   \end{eqnarray}
These equations have very clear physical meaning. The propagator in
the loop is
retarded (advanced), and guarantees the required time direction.
It is  affected
by the surroundings less than other correlators. The correlator $D_1$
describes the
density of states which develops in course of the evolution. Thus, even the
light-cone propagation is not really  free -- emission introduces
additional phase shifts which result in the assembly of special wave packet.

The sum and the difference of the
Eqs.~(\ref{eq:G23A}), respectively, are the equations for the
real ($\Pi_s$) and
imaginary ($\Pi_0 /2$) parts of the retarded self-energy:
\begin{eqnarray}
\Pi^{\mu\nu}_{s}(p)= ig_{r}^{2} \int  {d^4 k \over (2\pi)^4}
[V^{\mu\alpha\lambda}_{acf}(p,-k-p,k) D_{s}^{cc',\alpha\beta}(k+p)
V^{\nu\beta\sigma}_{bc'f'}(-p,p+k,-k)D_{1}^{f'f,\sigma\lambda}(k)~~,
\label{eq:G23B}\end{eqnarray}
\begin{eqnarray} \Pi^{\mu\nu}_{0}(p)= ig_{r}^{2} \int
{d^4 k \over (2\pi)^4} [ V^{\mu\alpha\lambda}_{acf}(p,-k-p,k)
D_{0}^{cc',\alpha\beta}(k+p)
V^{\nu\beta\sigma}_{bc'f'}(-p,p+k,-k)D_{1}^{f'f,\sigma\lambda}(k)~~.
\label{eq:G23C}\end{eqnarray}

Using Eqs.~(\ref{eq:G1}),~(\ref{eq:G17}), and (\ref{eq:G21}), and projecting
Eq.~(\ref{eq:G23B}) onto the transverse normal mode, we arrive at
\begin{eqnarray}
w_{1}^{s}(p)= w_{(0)1}^{s}(p)
+{Z_1g_{r}^{2}N_c \over 2(2\pi)^3} \left[ \int d^4 k \delta[(k-p)^2]
{\cal I}({p^+\over k^+}){w_{1}^{s}(k)+(1-Z_3)k^2 \over [k^2]^2}
+{i \over 2\pi}  \int d^4 k {{\cal P}\over (k-p)^2}
{\cal I}({p^+\over k^+}) {w_{1}^{1}(k)\over [k^2]^2} \right]
\label{eq:G30}
\end{eqnarray}
where ${\cal P}$ denotes principal value integration.
Another way to derive this equation is to separate real and imaginary
parts in
the Eq.~(\ref{eq:G29}).  This consistency is a consequence of the dispersion
relation for $\Pi_{ret}$ which allows one to recover  ~Re$\Pi_{ret}$  via the
already known ~Im$\Pi_{ret}$. It reassures us that we are considering the
propagation of the gluon field in  proper environment of the pre-collision
cascade.

One may easily see that the imaginary part of the self-energy, considered
given, defines the free term in  the inhomogeneous equation
(\ref{eq:G30}) for the real part of the gluon self-energy. The unusual
feature of this equation (which is common for all ladder-type equations
like (\ref{eq:G27})) is the counter-term which is  detached from the
vacuum  part of the self-energy. The latter is divergent, and the
corresponding  counter-term is in the integrand of the equation. This
immediately requires  that the kernel should act on the counter term as a
$\delta(k-p)$. Futhermore, the integral which contains the counterterm,
by inspection, is proportional to the
one-loop vacuum self-energy of a gluon.
After the UV-renormalization, the latter typically
behaves like $~p^2 log(p^2/\Lambda^2)$, where
{}~$\Lambda$ is the IR cut-off mass.
We find that in order to have the counter-term
$ -p^2(1 - Z_3)$ in its legitimate place near
$w_{(0)1}^{s}(p)$, the following relation should hold~:
$\alpha_r  \sim 1 / \log (p^{2}/\Lambda^2)$.

We have  estimated the contribution of the counterterm by
first calculating the imaginary part of the
retarded gluon self-energy, and then singling out the logarithmic terms
in the dispersion integral for the real part. We obtain
 \begin{equation}
\alpha_r(p_{t}^{2})={4\pi \over \beta_0\log(p_{t}^{2}/\Lambda^2)}~ ~ ~.
 \label{eq:G32}
\end{equation}
Thus, the known behavior of the running coupling constant is recovered.
Since we have used very rough approximations, the exact equality of the
coefficient deserves  further study. In the renormalization group approach
it is a direct consequence of the Slavnov-Taylor identity in the
null-plane gauge: $~Z_1=Z_3$.
After this renormalization the integral equation~(\ref{eq:G30})
for the real part of the gluon self-energy $[w_{1}^{s}(p)]^{ren}$
takes its final shape:
\begin{eqnarray}
w_{1}^{s}(p)= w_{(0)1}^{s}(p)
+{\alpha_{s}(p_{t}^{2})N_c \over (2\pi)^2}  \int d^4 k \delta[(k-p)^2]
{\cal I}({p^+\over k^+}){w_{1}^{s}(k) \over [k^2]^2}
+i{\alpha_{s}(p_{t}^{2})N_c \over (2\pi)^3} \int d^4 k {{\cal P}\over (k-p)^2}
{\cal I}({p^+\over k^+}) {w_{1}^{1}(k)\over [k^2]^2}~~,
\label{eq:G31}
\end{eqnarray}
where the superscript ``$ren$'' is omitted.

Despite the remaining uncertainty caused by the approximation,
it seems to be very important  that the running coupling has appeared as a
consequence of causal evolution. Only the processes which took place inside the
past light cone of the local interaction contribute to the magnitude of the
coupling in its vertex. This guarantees the proper balance between propagation
and emission-absorbtion processes  at the pre-collision stage.

    We still have freedom to chose $Z_3$. It has not yet been used in
the renormalization; the running coupling has appeared as a necessary
condition for renormalization rather than as an explicit choice of
some physical parameters at some given 4-momentum.
The strategy behind this choice must be the same as in the old-fashioned
on-mass-shell renormalization of the asymptotic state:  the on-mass-shellness
means that the field propagation is steady, despite background
vacuum fluctuations.  Selecting this kind of boundary condition, we
cannot describe the dynamics which  leads to the ``undressing''
of the quark as required by the resonant condition of deep inelastic
scattering. Moreover, the imaginary part of the self-energy must
equal zero at the renormalization point.

Thus, the object we now study is a field configuration which has quite
different properties that those of a free particle.
This is evident, for example, from Eq.~(\ref{eq:G15}),
which indicates that imaginary part of $w^{R}_{1}(p)$
is strongly peaked near $p^- = 0$. This configuration is singled out by two
requirements (boundary conditions): (i) at the end of its evolution,
it produces an off-shell quark  that can interact with the electron
in a resonant way; (ii) this  quark stays bare (on-shell)
after the scattering.

In e-p deep inelastic scattering the second condition does not look too
realistic. The bare quark will immediately fragment into a hadronic jet.  In
p-p collisions, the electron is replaced by a gluon from the second proton,
but the final quark state must still propagate in the physical vacuum
and decays into hadrons as well.  Only in AA-collisions do we expect
the creation of a volume of perturbative vacuum large enough to allow
almost stable free propagation.

The most important point is that the pre-collision dynamics of the field
fluctuations is
the same in all three cases. This is guaranteed by the geometry
of the high energy collision and the causality principle. However,
one should keep in mind
that the type of fluctuations studied by DIS are strictly selected by the
trigger of the specific measurement. Imposing other triggers will select
other types of fluctuations. It can not be ruled out {\it a priori}
that the pieces $w_2$ and $\sigma_2$ that were neglected in LLA may
become more significant.

To find an analogy between the pre-collision dynamics of the
proton constituents and physics of continuous media, we may try
associate components of gluon self-energy with electromagnetic susceptibility.
If its imaginary part is infinite then we have an ideally conducting medium.
(Remember that $p^-$ is a ``frequency'' corresponding to the ``time'' $x^+$ .)
A significant growth of the imaginary part of the self-energy at the
``resonance'' $p^-=0$ should lead to anomalous dispersion -- the real
part must drop to zero~:
 \begin{equation}
 {\rm Re} w^{R}_{1}(p_{t}^{2}, p^+, p^- =0)-(1 -Z_3)(-p_{t}^{2}) =0~ ~ ~.
 \label{eq:G33}
\end{equation}
(We can suggest the formal mathematical argument: once
{}~Im$\Pi_{ret}\sim \delta(p^-)$, then from dispersion relation
{}~Re$\Pi_{ret}\sim {\cal P}(1/p^-)$ which,
though in a singular manner, is equal
to zero at $p^-=0$.)
In the region of anomalous dispersion, the phase and the group velocities
should have the opposite signs.  This reveals one more unusual feature of the
``undressing''
process: the phases of the fields participating in the assembly
of the wave packet representing an interacting quark (or gluon) travel in
the ``normal'' time-direction (from the past to the future).
These fields leave ``holes'' in the sea.
Only the act of measurement (scattering) transforms the creation of
these ``holes'' into the process of multiple emission.

 \renewcommand{\theequation}{5.\arabic{equation}}
\setcounter{equation}{0}
\bigskip
{\bf \Large 5.~Distributions of quarks and gluons in leading order.}
\bigskip

In this and following Sections we present results of
an explicit calculation of the
single-particle distribution of light quarks and gluons produced at the
earliest
stage of an AA-collision.  In this Section we calculate cross-sections
to the lowest order.  It is general practice to associate the corresponding
processes with the excitation of the sea quarks and gluons. There are three
processes of this type. Their graphs are shown in Fig.\ref{Fig1}
In Section~6 we will derive a complete set of equations for the
first order processes, and discuss how one avoids difficulties which
accompany calculations based on the factorization theorem \cite{Fact}.

\bigskip
\bigskip
{\underline{\it   5.1.~Production of light quarks to leading order.}}
\bigskip

We return to the initial formulae (\ref{eq:E6}) and (\ref{eq:E8}), and
rewrite the former in the momentum representation~:
\begin{equation}
{ dN_q \over d{\bf p} d^4 x}
 ={{\rm Tr} [ i \not p   \Sigma_{01}^{ii}(p)] \over (2\pi)^3 2p^0}~ ~ ~.
\label{eq:H1}
\end{equation}
The lowest order of our theory assumes: (i) the exact vertex operator
must be replaced by the bare one, which leads to
\begin{equation}
p^0 { dN_q \over d{\bf p}d^4 x}
 ={g^2 \over 2(2\pi)^3} \int {d^4 k \over (2\pi)^4 }
  {\rm Tr} [\not p  t^a \gamma^{\mu} {\bf G}_{01}(p+k)t^b \gamma^{\mu}
{\bf D}^{ba}_{10,\nu\mu}(k) ]~ ~ ~;
\label{eq:H2}
\end{equation}
and (ii) in ${\bf G}_{01}$ and ${\bf D}_{10}$ possible contribution of the
out-states of quark and gluon in the continuum are excluded.
These contributions are
described by the higher orders of the perturbation theory. Thus, we naturally
arrive at Eq.~(\ref{eq:E8})
\begin{equation}
p^0 { dN_q \over d{\bf p}d^4 x} ={g^2 \over 2(2\pi)^3}
\int {d^4 kd^4 q \over (2\pi)^4 }\delta(k+q-p)
\{ {\rm Tr} [\not p  t^a \gamma^{\mu} {\bf G}^{(B)}_{01}(q)t^b
\gamma^{\mu}
{\bf D}^{(A)ba}_{01,\nu\mu}(k) ] + (A) \leftrightarrow (B) \}~ ~ ~,
\label{eq:H3}
\end{equation}
were the quark and gluon correlators must be taken in the following form:
\begin{eqnarray}
{\bf G}^{(J)}_{01}= G^{*(J)}_{01}-{\bf G}^{(J)}_{ret}\Sigma^{(J)}_{01}
{\bf G}^{(J)}_{adv}, \;\;\;\;\;
{\bf D}^{(J)}_{01}= D^{*(J)}_{01}-{\bf D}^{(J)}_{ret}\Pi^{(J)}_{01}
{\bf D}^{(J)}_{adv},\;\;\;\;\;(J=A,B),
\label{eq:H4}
\end{eqnarray}
with $G^{*(J)}_{01}$ and $D^{*(J)}_{01}$ representing the quark and gluon
distributions at some arbitrary scale $Q^{2}_{0}$. For computations,
we shall use a standard CTEQ parameterization of the nucleon structure
functions \cite{CTEQ}  $q(x, q_{t}^{2})$ and $G(x, k_{t}^{2})$. These were
obtained by fitting the data with the solutions of the GLAP evolution
equations. For nuclei, we shall also use the semi-empirical
formula fit to nuclear shadowing data.

We proceed in the laboratory frame, which is the infinite momentum
frame for both
proton A and proton B. The directions of their light cone propagation are
fixed by the two null-plane vectors $n^{\mu}_{A}$ and $n^{\mu}_{B}$:
\begin{eqnarray}
 n^{\mu}_{A}= (1,{\bf 0}_t, -1),\;\;\;\;n^{\mu}_{B}= (1,{\bf 0}_t, 1),
\;\;\;\;\;   n^{2}_{A}= n^{2}_{B}= 0.
\label{eq:H5}\end{eqnarray}
They define the light-cone components of the Lorentz vectors:
\begin{eqnarray}
 n_Aa = a^+ = a_- = a^0 + a^3,\;\;\;\;  n_B b = b^- = b_+ = b^0 - b^3.\nonumber
\end{eqnarray}

    We split the total cross-section into the three parts:
\begin{eqnarray}
 \sigma^{(0)}_{q}=\sigma^{(0)}_{q}({\cal V},\Pi)+
\sigma^{(0)}_{q}({\cal G},\Sigma)+ \sigma^{(0)}_{q}(\Pi,\Sigma),
\label{eq:H6}
\end{eqnarray}
The term $\sigma^{(0)}_{q}({\cal G},{\cal G})$ is naturally absent, as
energy-momentum conservation prevents fusion of the two on-mass-shell
particles into one on-mass-shell quark.

  Unlike the case of DIS,
both invariants from quark ($\sigma_1$ and $\sigma_2$)
and gluon ($w_1$ and $w_2$) self-energies contribute to the cross-section of
quark production. In order not to exceed the accuracy of the leading
logarithmic approximation of the GLAP-evolution of structure functions, we
shall omit $\sigma_2$ and $w_2$. They are of the next order by a formal
count of the $\alpha_s$-powers, and are not under direct control of DIS data.

Any term coming from the nucleus A carries a $\delta(k^-)$, and any term coming
from the nuclei B carries a $\delta(q^+)$. This drastically simplifies the
calculation. The first two  terms from Eq.~(\ref{eq:H6}) are calculated
explicitly, yielding
\begin{eqnarray}
{d \sigma^{(0)}_{q}({\cal V},\Pi)\over  dp_{t}^{2}dy }=
{16\pi^2\alpha_0 \over 3s}[q({p_te^{-y}\over \sqrt{s}},Q_{0}^{2})
G'({p_te^{y}\over \sqrt{s}},p_{t}^{2})+(y\rightarrow -y)],
\label{eq:H7}
\end{eqnarray}
\begin{eqnarray}
{d \sigma^{(0)}_{q}({\cal G},\Sigma)\over  dp_{t}^{2}dy }=
{16\pi^2\alpha_0 \over 3s}[G({p_te^{-y}\over \sqrt{s}},Q_{0}^{2})
q'({p_te^{y}\over \sqrt{s}},p_{t}^{2})[1+{p_{t}^{2}\over sx_Ax_B}]
+(y\rightarrow -y)]~ ~ ~,
\label{eq:H8}
\end{eqnarray}
where $y$ is the longitudinal rapidity of the final state quark,
$x_{A,B}=p_te^{\pm y}/\sqrt{s}$, and the substitute term
$(y \rightarrow -y)$ accounts for the symmetry $(A \leftrightarrow B)$.
We have also defined  $G'(x,p_{t}^{2})=d G(x,p_{t}^{2})/d p_{t}^{2}$ and
$q'(x,p_{t}^{2})=dq(x,p_{t}^{2})/d p_{t}^{2}$.
The main contribution to the cross-section
comes from the third term in Eq.~(\ref{eq:H6}):
\begin{eqnarray}
{d \sigma^{(0)}_{q}(\Sigma,\Pi)\over  dp_{t}^{2}dy }=
{8\pi\alpha_0 \over 3sp_{t}^{2}}
\int {dk_{t}^{2} dq_{t}^{2}(k_{t}^{2}+q_{t}^{2})(1+q_{t}^{2}/sx_Ax_B) \over
\sqrt{[(k_{t}+q_{t})^{2}-p_{t}^{2}]
[p_{t}^{2}-(k_{t}-q_{t})^{2}]} } [q'({ p_te^{-y}\over \sqrt{s} },q_{t}^{2})
G'({ p_t e^{y} \over \sqrt{s} },k_{t}^{2})+(y \rightarrow -y)]~ ~ ~,
\label{eq:H9}
\end{eqnarray}
where all terms which are at least as small as the
longitudinal structure function
$F_L$ of the DIS were neglected.
The square root in the denominator comes from the angular integration over the
orientations of the transverse momenta:
\begin{eqnarray}
\int  d^{2}{\bf k}_{t} d^{2}{\bf q}_{t}
\delta^{(2)}({\bf k}_{t}+{\bf q}_{t}-{\bf p}_{t})F(k_t,q_t) =
\int { dk_{t}^{2}dq_{t}^{2} F(k_t,q_t)\over 4S(k_t,q_t,p_t) }~'
\label{eq:H10}
\end{eqnarray}
where $S(k_t,q_t,p_t)$ is the area of the triangle with  sides
$k_t$, $q_t$ and $p_t$,
\begin{eqnarray}
 4S(k_t,q_t,p_t)=\sqrt{[(k_{t}+q_{t})^{2}-p_{t}^{2}]
[p_{t}^{2}-(k_{t}-q_{t})^{2}]}   ~ ~ ~,
\label{eq:H11}
\end{eqnarray}
and the integration domain is restricted by the triangle inequalities,
\begin{eqnarray}
 k_{t}+q_{t}\geq p_{t},\;\;\;\; |k_{t}-q_{t}|\leq p_{t}~ ~ ~.
\label{eq:H12}
\end{eqnarray}

\bigskip
\bigskip
{\underline{\it   5.2.~Production of gluons to leading order.}}
\bigskip

For the case of gluon production, we start from
Eq.~(\ref{eq:E9}), and, as for Eq.~(\ref{eq:H1}),
rewrite it in the momentum representation~:
\begin{equation}
{ dN_g \over d{\bf p}d^4 x}
 ={d_{\mu\nu}(p,u)[- i  \Pi_{01}^{aa,\mu\nu}(p)] \over (2\pi)^3 2p^0}
\label{eq:H13}
\end{equation}
where the projector $d_{\mu\nu}(p,u)$
\begin{equation}
d^{\mu\nu}(p,u)=-d_{\mu\nu}+{p^{\mu}u^{\nu}+u^{\mu}p^{\nu} \over (pu)}
-{p^{\mu}p^{\nu} \over (pu)^2},\;\;\;\;
d^{\mu\nu}(p,u)u_{\nu}=0,\;\;\;\; p^2=0~ ~ ~,
\label{eq:H14}
\end{equation}
is a sum over the two physical gluon polarizations in  laboratory frame with
time axis along the four-vector $u^{\mu} = (1,0,0, 0)$. For this gluon
we choose the time-like axial gauge $u^{\mu}B^{a}_{\mu}= 0,\;\;  u^2 = 1$.

To the same approximation as in Eq.~(\ref{eq:H2}), we may write the gluon
polarization tensor (temporarily omitting the fermion contribution) as
\begin{eqnarray}
  \Pi^{\mu\nu}_{01}(p)=-ig_{0}^{2}
\{\! \int \!\!  {d^4 kd^4 q \over (2\pi)^4} \delta(k + q - p)
 \{ V^{\mu\rho\sigma}_{acd}(k+q,-q,-k)
{\bf D}_{01,dd'}^{(A)\sigma\beta}(k)
 V^{\nu\lambda\beta}_{bc'd'}(-k-q,q,k)
 {\bf D}_{10, f'f}^{(B)\rho\lambda}(q)]+
(A \leftrightarrow B) \}
\label{eq:H15}
\end{eqnarray}
The gluon distribution of each nucleus consists of the two familiar terms
given by Eq.~(\ref{eq:H4}).  Again, we shall keep
only the leading terms in the gluon sources which are under the
control of the data~:
\begin{equation}
[d(k,n_A)\Pi^{(A)}(k)d(k,n_A)]^{\mu\nu}_{dd'}\approx -{\delta^{dd'}\over 8}
{\bar d}^{\mu\nu}(k,n_A) w_{1}^{A}(k)~ ~ ~.
\label{eq:H16}
\end{equation}
A similar expression may be written down for nucleus B.
Calculation of the trace over the abundant
color and vector indices results in an overall factor of
$24\times 32\times T(k,q)$.
The trace $T (k,q)$ is given below. As in Eq.~(\ref{eq:H6})
we have two types of contribution to the total cross-section~:
\begin{eqnarray}
 \sigma^{(0)}_{g}=\sigma^{(0)}_{q}({\cal G},\Pi)+
 \sigma^{(0)}_{q}(\Pi,\Pi),
\label{eq:H17}
\end{eqnarray}
The first term comes from the interaction of the ``source''
with the field from the ``initial'' distribution~:
\begin{eqnarray}
{d \sigma^{(0)}_{g}({\cal G},\Pi)\over  dp_{t}^{2}dy }=
{12\pi^2\alpha_0 \over s }[G({p_te^{-y}\over \sqrt{s}},Q_{0}^{2})
G'({p_te^{y}\over \sqrt{s}},p_{t}^{2})
\left[1+{p_{t}^{2} e^{2y} \over s (x_A+x_B)^2}\right]+(y\rightarrow
-y)]~ ~ ~.
\label{eq:H18}\end{eqnarray}
The second term comes from the interaction of the two sources~:
\begin{eqnarray}
{d \sigma^{(0)}_{g}(\Pi,\Pi)\over  dp_{t}^{2}dy }=
{24\pi\alpha_0 \over sp_{t}^{2}}
\int { dk_{t}^{2}dq_{t}^{2}T(k_{t},q_{t}) \over 4S(k_t,q_t,p_t)}
G'({p_te^{-y}\over \sqrt{s}},k_{t}^{2})
G'({p_te^{y}\over \sqrt{s}},q_{t}^{2})~ ~ ~.
\label{eq:H19}
\end{eqnarray}
The trace $T(k,q)$ is conveniently written in the following way~:
\begin{eqnarray}
 T(k,q) = k_{t}^{2}+q_{t}^{2} +{(k_{t}^{2}+q_{t}^{2})^2 \over s(x_A+x_B)^2}
\left[{q_{t}^{2} \over sx_{B}^{2}}+ {k_{t}^{2} \over sx_{A}^{2}}+
{k_{t}^{2}q_{t}^{2} \over s^2 x_A x_B}\right]+
{2k_{t}^{2}q_{t}^{2} \over s(x_A+x_B)^2}
\left[{q_{t}^{2} \over sx_{A}^{2}}+ {k_{t}^{2} \over
sx_{B}^{2}}\right]~ ~ ~.
\label{eq:H20}\end{eqnarray}
The integrand of Eq.~(\ref{eq:H19}) is symmetric with respect to interchange
of $k_t$ and $q_t$. Hence
the contribution from the interchange term $(A \leftrightarrow B)$
is the same, and factor 2 is already included in Eq.~(\ref{eq:H19}).

We close this Section by discussing two minor problems. The first
is an apparent infrared divergence of total cross-section due to the
kinematic factor $p_{t}^{-2}$ in Eqs. ~(\ref{eq:H9}) and ~(\ref{eq:H19}).
However, one can easily see that this factor disappears when the integration
variables $k_t$ and $q_t$ are rescaled by $p_t$. After such a
rescaling, any possibly remaining issues concerning
low $p_t$ behavior are connected with the sources (or structure
functions) which are controlled by the data.

The second problem arises from
the positive powers of transverse momenta of the incoming quark and
gluon fields. This can be seen from Eqs.~(\ref{eq:H9}), and
(\ref{eq:H19}) and (\ref{eq:H20}), and appear in the same  form
in the next perturbation orders. The formal solution of the problem is
very simple: to be consistent with the condition
under which the sources (or structure functions) are
defined, {\it viz.}, $s \rightarrow \infty $,
we must drop all terms containing ratios like
$q_{t}^{2}/sx_Ax_B$  and $k_{t}^{2}/sx_Ax_B$, despite the external
kinematic condition $sx_Ax_B=p_{t}^{2}$.
If this is not done then the integrals  (\ref{eq:H9}) and
(\ref{eq:H19}) will diverge,
as the LLA structure functions do not provide sufficient cut-off
at high $k_t$ and $q_t$.  In the context of
the existing theory, these ratios are parametrically small and may
not be switched on until the high-order twists are taken
into account.

Nevertheless, the growth of the matrix element at high transverse momenta
is quite physical, and may be interesting at very small $x$.
The wave packets with small $x$ are smooth and
extended in the longitudinal direction. If we require that large
$k_t$ and  $q_t$
add  to form small $p_t$, then the initial geometry of momenta
is almost collinear and the incoming fields  effectively overlap.
We thus encounter a type of collinear singularity which may be shielded only
by appropriate behavior of structure functions.
The latter is naturally
provided by the BFKL equation (See Ref.\cite{BFKL} and Appendix 1.)
We emphasize that this process may be taken seriously only in
the domain of very low $x$, where only very preliminary data exist at the
present time.

 \renewcommand{\theequation}{6.\arabic{equation}}
\setcounter{equation}{0}
\bigskip
{\bf \Large 6.~First order corrections to the quark distribution.}
\bigskip

In the previous Section we have calculated the lowest (Born)
contribution  (of $\alpha_{s}^{0}$-order)
to  quark and gluon production corresponding to the ``excitation'' of a single
quark or gluon from the sea.  Now we intend to show that the
next order terms of our kinetic perturbation expansion describe the
more usual ``creation''
processes. They do it in a very special way such that no infrared
divergencies appear in the calculation of the total cross-section.
This may be the most important result of the present calculation.
Furthermore, this requires no special proof: because the perturbation
series for the {\it probabilities} was initially resummed
in the new expansion, it does not generate any terms
affected by the initial state collinear singularities. These
are absorbed into definition of the sources (structure functions),
and thus solving the problem of divergence in an
experimental way.

This is in striking
contrast to all known calculations based on the factorization theorem.
Such calculations always start from a {\it master formula}, such as
\begin{eqnarray}
p^0 { d\sigma(AB\rightarrow pX) \over d{\bf p}}
 =\sum_{a,b}\sum_{c\in X} \int_{0}^{1}dx_a \int_{0}^{1}dx_b
 p^0 { d\sigma(a,b\rightarrow p,c) \over d{\bf p}}
F_{aA}(x_a,Q^2) F_{bB}(x_b,Q^2)~ ,
\label{eq:N0}
\end{eqnarray}
which heavily relies on the parton model, and implicitly impart the status of
an observable at least to  one additional final state particle.
The $2\rightarrow 2$ processes
have the lowest order in this approach, and the $Q^2$-dependence of
the structure
functions $F_{jJ}$ defines the so-called factorization scale, rather than
reflecting its full QCD-evolution.  As was already mentioned in Section~4, the
reason structure functions are treated mistrustfully in this approach
is connected with
the out-of-light-cone behavior of Feynman's propagators for the
incoming partons.

Quite naturally, apart from all other amplitudes, the $2\rightarrow 2$
cross-section includes those corresponding
to the emission of a second particle $c\in X$ from the initial state.
The squared moduli of these partial amplitudes duplicate those already included
in the definition of the structure functions at lower factorization scales.

The new approach does not distinguish any states absorbed into the set $X$.
At the same time, it restores the proper status of the retarded
propagation for the incoming fields. As a result, it
excludes {\it a priori} any double counting of processes.

    The first order corrections to the quark or gluon production can be divided
into two major categories: the self-energy-like and the
vertex-like. These names reflect only the topology of the new diagrams.

\bigskip
\bigskip
{\underline{\it 6.1.~Self-energy-like terms.}}
\bigskip

The sequence of the $\alpha_{s}$-order diagrams in the
self-energy-type term emerges from
the possibility that either the quark of gluon field correlator
in Eq.~(\ref{eq:H2}) represents a field in the continuum of
out-states.
\begin{eqnarray}
p^0 { dN_q \over d{\bf p}d^4 x}
 ={g_{0}^{2} \over 2(2\pi)^3} \int {d^4 k \over (2\pi)^4 }
  \{ {\rm Tr} [\not p t^a \gamma^{\mu}  G^{\#}_{01}(p+k)t^b \gamma^{\mu}
{\bf D}^{ba}_{10,\nu\mu}(k) ]
+ {\rm Tr} [\not p t^a \gamma^{\mu} {\bf G}_{01}(p+k)t^b \gamma^{\mu}
 D^{\# ba}_{10,\nu\mu}(k) ] \},
\label{eq:N1}
\end{eqnarray}
This immediately means that the remaining exact correlators ${\bf G}_{01}$
and ${\bf D}_{10}$ carry information about both nucleus A and nucleus
B. In other
words, these fields were created during the course of the collision
between the two nuclei.
The only terms of the subsequent expansion of the quark and  gluon
correlator that survive in this case are
\begin{eqnarray}
{\bf D}_{10}=  - {\bf D}^{\#}_{ret}\Pi_{10} {\bf D}^{\#}_{adv},\;\;\;\;
{\rm and} \;\;\;
{\bf G}_{01}=  - {\bf G}^{\#}_{ret}\Sigma_{01} {\bf G}^{\#}_{adv}
\label{eq:N3}
\end{eqnarray}
The superscript ``$\#$''
indicates that we consider propagation after collision of the
nuclei, and in case we need radiative corrections to this propagation, we must
consider them against the background of the distribution of quarks and
gluons created in the collision itself.

The intensities $\Pi$ and $\Sigma$ of the field sources created by
the two nuclei, were already calculated in the previous Section, but
now these fields are
off-mass-shell, and the terms with products of two $\ast$-labelled
correlators from equations  (\ref{eq:H4}) should be added.
These are just the lowest order terms of the master formula (\ref{eq:N0})
of the QCD
parton model, with a fixed factorization scale  $Q_{0}^{2}$, and
correspond to processes of the type $2\rightarrow 2$, where
parameters of the second emitted particle are completely integrated over.
Thus the complete formula will be
\begin{eqnarray}
p^0 { dN_q \over d{\bf p}d^4 x}
 ={-g_{0}^{2} \over 2(2\pi)^3} \int {d^4 k d^4 q \over (2\pi)^4 }
  \{ Tr [\not p t^a \gamma^{\mu}  G^{\#}_{01}(p+k)t^b \gamma^{\mu}
[{\bf D}^{\#}_{ret}(k)\Pi^{AB}_{10}(k)
{\bf D}^{\#}_{adv}(k)]^{ba}_{\nu\mu}] +  \nonumber  \\
+ Tr [\not p t^a \gamma^{\mu}
 [{\bf G}^{\#}_{ret}(k)\Sigma^{AB}_{01}(k) {\bf G}^{\#}_{adv}(k)]t^b
\gamma^{\mu}
 D^{\# ba}_{10,\nu\mu}(k-p) ] \}~ ~ ~,\hspace*{0.5cm}
\label{eq:N4}
\end{eqnarray}
where the following chain of substitutions is supposed~:
\begin{eqnarray}
\Sigma^{AB}_{01}(p)
 =ig_{0}^{2}\int {d^4 k \over (2\pi)^4 }  \delta (k+q-p)\{ t^a \gamma^{\mu}
[ G^{B*}_{01}(k)-{\bf G}_{ret}(k)\Sigma^{B}_{01}(k) {\bf G}_{adv}(k) ]
t^b \gamma^{\mu} \times \nonumber    \\
( D^{A*}_{10}(q)- {\bf D}_{ret}(q)\Pi^{A}_{10}(q){\bf D}_{adv}(q) )_{ba,\nu\mu}
  + (A \leftrightarrow B)\}\hspace*{1.5cm}
\label{eq:N5}
\end{eqnarray}
for a ``collective'' quark source, and
\begin{eqnarray}
\Pi^{AB}_{01}(p)
 =-ig_{0}^{2} \int {d^4 k d^4 q \over (2\pi)^4 }\{ ~V~
[D^{B*}_{10}(k)- {\bf D}_{ret}(k)\Pi^{B}_{10}(k)
{\bf D}_{adv}(k)] ~VV [ D^{A*}_{10}(q)-{\bf D}_{ret}(q)
\Pi^{A}_{10}(q){\bf D}_{adv}(q) ]+ \nonumber\\
+{\rm Tr} t\gamma[ G^{B*}_{01}(k)-{\bf G}_{ret}(k)
\Sigma^{B}_{01}(k){\bf G}_{adv}(k) ]
 t\gamma[ G^{A*}_{10}(q)-{\bf G}_{ret}(q)
\Sigma^{A}_{10}(q){\bf G}_{adv}(q) ] \} \delta (k+q-p)~  ~ ~,
\label{eq:N6}
\end{eqnarray}
for the ``collective'' gluon source.
Here $V$-s stand for the 3-gluon vertex with all arguments dropped.

Three of the six graphs of this type are given in Fig.\ref{Fig2}. The first
corresponds to the s-channel part of the Compton process.  The other
two are of an
annihilation type.  They can all be represented as squared moduli of the
corresponding amplitudes.
The t-channel partners of these diagrams are not generated
by this perturbative expansion. If they did, they would duplicate
processes already included in the definition of the sources. The
three omitted graphs correspond to the interchange ($A \leftrightarrow B$).

\bigskip
\bigskip
{\underline{\it 6.2.~First order terms from the ``vertex-like'' corrections.}}
\bigskip

The general expressions for the one-particle quark and gluon distributions
(Eqs.~(\ref{eq:E8}) and (\ref{eq:E11})) contain dressed quark-gluon and
three-gluon vertices
\begin{eqnarray}
 \Gamma^{d,\lambda}_{SQ,P}(\xi,y;\eta)=(-1)^{P+S+Q}
{ {\delta [{\bf G}^{-1}(\xi,y)]_{SQ} } \over
 {g_{r} \delta {\cal B}^{d}_{\lambda}(\eta_{P}) }  }   , \nonumber \\
   {\bf V}^{\nu\beta\sigma}_{bcf,RSB}(\xi,\eta,y)=(-1)^{R+S+P}
{ {\delta [{\bf D}^{-1}(\xi,\eta)]^{bc;\nu\beta}_{RS} } \over
  {g_r \delta {\cal B}^{f}_{\sigma}(y_{B}) }  } .
\label{eq:N7}
\end{eqnarray}
Because of their additional matrix structure they contain some unusual elements
which should be determined now.  It can be done easily by using the formal
solution of the matrix Schwinger-Dyson equations~:
\begin{equation}
 [{\bf G}^{-1}]_{AB}  =  [G^{-1}]_{AB}- \Sigma_{AB}, \;\;\;\;
 [{\bf D}^{-1}]_{AB}  = [ D^{-1}]_{AB}- \Pi_{AB}.
\label{eq:N8}
\end{equation}
Functional derivatives of the bare Greenians give us the bare vertices, and
those of self-energies give the corrections. Using the one-loop formulae for
the self-energies it is straightforward to find the first order corrections to
the vertices~:
\begin{eqnarray}
^{(1)}\Gamma^{d,\lambda}_{RB,S}(\xi,y;\eta)= -i g_{r}^{2}(-1)^{R+S}\{
\gamma^{\rho}{\bf t}^a {\bf G}_{RS}(\xi,\zeta)\gamma^{\lambda}{\bf t}^d
{\bf G}_{SB}(\zeta ,y)\gamma^{\sigma}{\bf t}^b
{\bf D}^{ba;\sigma\rho}_{SR} (y, \xi)+ \nonumber  \\
+\gamma^{\rho}{\bf t}^a {\bf G}_{RB}(\xi,y)\gamma^{\sigma}{\bf t}^b
{\bf D}^{bc;\sigma\beta}_{BS} (y, \zeta) V_{\alpha\beta\lambda}^{cgd}(\zeta)
{\bf D}^{bc;\alpha\rho}_{BS}(\zeta,\xi) \}~ ~ ~.  \hspace*{1cm}
\label{eq:N9}
\end{eqnarray}
Eq.~(\ref{eq:N9}) contains an additional rule: besides the expected
contour indices, the correction $^{(1)}\Gamma^{d,\lambda}_{RB,S}(\xi,y;\eta)$
to the bare vertex with the contour index $B$ acquires an additional factor
$(-1)^{R+S}$. This rule works in the same way for the three-gluon vertex which
is too cumbersome to be written down separately.

    The sum over the contour indices naturally divides into two groups: those
with $R=S$ and $R\neq S$.  We delay discussion of the first group which is
responsible for those types of
radiative corrections which may lead to vertex
form-factors.  The second group, which we will discuss below, describes
contributions of the higher-order real processes, like emission of a second
uncontrolled jet.

The one-loop vertex-type correction to single light quark production contains
four terms, two from the first term in Eq.~(\ref{eq:N9}) and two from
the second one~:
\begin{eqnarray}
p^0 { dN_q \over d{\bf p}d^4 x} ={-ig_{0}^{2} \over 2(2\pi)^3}
\sum_{j=1}^{4}\int {d^4 k d^4 q \over (2\pi)^8 }
{\rm Tr}[\not \! p {\cal I}_{j}(p,k,q)],
\label{eq:N10}
\end{eqnarray}
where
\begin{eqnarray}
{\cal I}_{1}&=&{\bf t}^a \gamma^{\mu} {\bf G}_{01}(k){\bf t}^d \gamma^{\sigma}
 {\bf G}_{10}(k+q-p){\bf t}^c \gamma^{\rho}
{\bf G}_{01}(q){\bf t}^b \gamma^{\nu}{\bf D}^{ac}_{00,\mu\rho}(p-k)
{\bf D}^{db}_{11,\sigma\nu}(p-q);  \nonumber  \\
{\cal I}_{2}&=&{\bf t}^a \gamma^{\mu} {\bf G}_{00}(p-k){\bf t}^d
\gamma^{\sigma}
 {\bf G}_{01}(p-k-q){\bf t}^c \gamma^{\rho}
{\bf G}_{11}(p-q){\bf t}^b \gamma^{\nu}{\bf D}^{ac}_{01,\mu\rho}(k)
{\bf D}^{db}_{01,\sigma\nu}(q);  \nonumber  \\
{\cal I}_{3}&=&{\bf t}^a \gamma^{\mu} {\bf G}_{00}(p-k){\bf t}^c
\gamma^{\sigma}
 {\bf G}_{01}(q){\bf t}^b \gamma^{\nu}
{\bf D}^{aa'}_{01,\mu\rho}(k){\bf D}^{cc'}_{01,\lambda\sigma}(k+q-p)
{\bf D}^{b'b}_{11,\phi\nu}(p-q)
 V^{\rho\lambda\phi}_{a'b'c'}(-k,k+q-p,p-q);\nonumber \\
{\cal I}_{4}&=&{\bf t}^a \gamma^{\mu} {\bf G}_{01}(k){\bf t}^c \gamma^{\sigma}
 {\bf G}_{11}(p-q){\bf t}^b \gamma^{\nu}
{\bf D}^{aa'}_{00,\mu\rho}(p-k){\bf D}^{cc'}_{01,\lambda\sigma}(p-k-q)
{\bf D}^{b'b}_{01,\phi\nu}(q)
 V^{\rho\lambda\phi}_{a'b'c'}(k-p,k+q-p,q)~~.
\label{eq:N11}
\end{eqnarray}
These four terms are depicted in Fig.\ref{Fig3}.
There are three cut lines in each diagram,
and they can be ``distributed'' in six different ways between the two colliding
nuclei and the additional out-state excited in the continuum. After that we can
convert
every term of this expansion into a product of two scattering amplitudes.
This is done in Fig.\ref{Fig4} (up to a trivial interchange
$A\leftrightarrow B$).
It is clearly seen that amongst this group of graphs there are none which
would represent the squared moduli of any
amplitude, but all allowed interference terms
between all the processes which produce a quark and something else are
included.

Recalling the previous discussion of the self-energy-type terms, we see that
our perturbative expansion does not generate any diagrams which would repeat
those present in the definition of the sources (via their ladder
expansion).  These missing patterns are not IR-safe, and were regularized and
renormalized in the course of their definition via the DIS cross-section.

We shall now show that no further infrared problems appear.
We demonstrate this using the definite subprocess of a detected quark
with momentum $p$ accompanied by an uncontrolled anti-quark in the out
state. They were both created in a collision of two gluons,
$g_Ag_B\rightarrow q{\bar q}$.

\bigskip
\bigskip
{\underline{\it 6.3.~Infrared safety in the $\alpha_s$-order.}}
\bigskip

Infrared finiteness of the self-energy-type diagrams of Fig.\ref{Fig2}
is intuitively
understandable, since the intermediate $s$-channel gluon or quark carry a
large time-like
momentum.  We may expect IR-problems only in the vertex-type diagrams,
like (V1) of Fig.\ref{Fig4}, where the
intermediate fermion is in the $t$-channel. The corresponding distribution of
a single quark is  given by the expression~:
\begin{eqnarray}
p^0 { dN_q \over d{\bf p}d^4 x} ={-ig_{0}^{2} \over 2(2\pi)^3}
 \int {d^4 k d^4 q \over (2\pi)^8 }
{\rm Tr}[\not p {\bf t}^a \gamma^{\mu}
{\bf G}_{00}(p-k){\bf t}^d\gamma^{\sigma}
 {\bf G}_{01}(p-k-q){\bf t}^c \gamma^{\rho}
{\bf G}_{11}(p-q){\bf t}^b \gamma^{\nu} ]  \times  \nonumber \\
\times \left[ D^{(A)*}_{01}(k)-{\bf D}^{(A)}_{ret}(k)
\Pi^{(A)}_{10}(k){\bf D}^{(A)}_{adv}(k)\right]^{ac}_{\rho\mu}
\left[ D^{(B)*}_{10}(q)-{\bf D}^{(B)}_{ret}(q)\Pi^{(B)}_{10}(q)
{\bf D}^{(B)}_{adv}(q)\right]^{bd}_{\sigma\nu}~~~,
\label{eq:N12}
\end{eqnarray}
where retarded and advanced functions carry a superscript indicating
which nucleus was the source of field.
The natural variables for subsequent calculations are quark rapidity and
momentum fractions defined via
\begin{eqnarray}
p^{\pm}=p_t e^{\pm y},\;\;\;\;\;k^+=\sqrt{s}x_A,\;\;q^-=\sqrt{s}x_B~ ~
{}~. \nonumber
\end{eqnarray}
The most troublesome element in all following calculations is the
trace over spinor and vector indices,
\begin{eqnarray}
T = {\rm Tr}[\not p \gamma^{\mu}(\not p- \not k)
\gamma^{\sigma}(\not p- \not k- \not q)\gamma^{\rho}(\not p- \not q)
\gamma^{\nu}]d_{\rho\mu}(k,n_A)d_{\sigma\nu}(q,n_B)~.
\label{eq:N13}
\end{eqnarray}

As in the lowest order, we can single out different types of terms
contributing to the cross-section in the first order:
\begin{eqnarray}
 \sigma^{(1)}_{q}=\sigma^{(1)}_{q}({\cal G},{\cal G})+
\sigma^{(1)}_{q}({\cal G},\Pi)+ \sigma^{(1)}_{q}(\Pi,\Pi)~,
\label{eq:N14}
\end{eqnarray}
The first term corresponds to the first nonvanishing
order of the master formula of the parton model, which factorizes the ``hard''
QCD cross-section and structure functions evaluated at some
(sufficiently high)
scale $Q^{2}_{0}$. Two mass-shell delta-functions make the
calculations relatively simple, and the result reads as follows:
\begin{eqnarray}
 {d\sigma^{(1)}_{q}({\cal G},{\cal G}) \over d p_{t}^{2} dy} =
-{\pi \alpha_{0}^{2} \over 12 s^2} \int_{0}^{1}
{dx_A dx_B \over x_{A}x_{B} } \left[1-{8p_{t}^{2}\over sx_Ax_B} \right]
  G(x_A,Q_{0}^{2})G(x_A,Q_{0}^{2}) \times  \nonumber   \\
\times\delta [x_Ax_B-{p_{t}\over \sqrt{s}}({x_Ae^{-y} +x_B}e^{y})]
\theta (x_A-{p_{t}\over \sqrt{s}}e^{y})\theta (x_B-{p_{t}\over
\sqrt{s}}e^{-y})~.
\label{eq:N15}
\end{eqnarray}
We see that this term remains finite when  $p_t \rightarrow 0$
and is strongly suppressed, both by the second power of $s$ in the denominator
and the smallness of $\alpha_{0}^{2}$.

To facilitate the following analysis, let us trace how the structure
emerges. We may
expect an infrared divergence from the poles of the product of the
two $t$-channel
propagators  in Eq.~(\ref{eq:N12}), $G_{00}(p - k)G_{11}(p - q)$.
Since they are
unshielded by finite masses or virtualities, the factor
$(p^{2}_{t}sx_Ax_B )^{-1}$ appears.
One power of $s^{-1}$ comes from the definition of
cross-section, and one more from the final-state phase space. The combined
spinor-vector trace in (\ref{eq:N14}) gives a factor $p^{2}_{t}s$.
As a result, no
large logarithm which could partially compensate for the
smallness of the coupling constant
appears in the total cross-section.
Eventually, we may expect only a weak scale dependence of
the total cross-section.

The ``box diagram'', which is an unavoidable partner of
$\sigma^{(1)}_{q}({\cal G},{\cal G})$  in the
approach based on the factorization theorem, is IR-divergent, but it
has no analog in our perturbation
expansion.   Nevertheless, we considered it useful
to calculate it in Appendix~2, in order to compare its structure with that
emerging from the present calculations.

The mixed term $\sigma^{(1)}_{q}({\cal G},\Pi)$,
and the term $\sigma^{(1)}_{q}(\Pi,\Pi)$ (contributed
to by two
``sources'') are more complicated because one or both of the
incoming fields interact
with their sources and hence are off the mass shell. This means that at least
one of the mass-shell delta functions is no longer present,
and traces become unwieldy.
We will not present their explicit form here, as these terms are not expected
to be large.  Indeed, after the box
diagram has been extracted, one retains only interference terms. These are
usually small, unless they are affected by a singular infrared behavior.
A sufficient qualitative analysis of this behavior can
be done  without explicit calculations. It is enough to notice that the trace
$T (p,k,q)$ in the numerator of the integrand of Eq.~(\ref{eq:N12}) is,
in general, a polynomial of fourth order with respect to $p_t/\sqrt{s}$.
When the momenta of both structure functions  are put on mass shell, the only
 powers that survive are $p^{2}_{t}/s$ and $p^{4}_{t}/s^2$. This leads
to an
immediate cancellation of the two unshielded infrared poles of the
propagators. In
the integral for $\sigma^{(1)}_{q}(\Pi,\Pi)$ both poles are shielded by gluon
virtualities, and infrared divergence can not appear at all. In the integral
 for $\sigma^{(1)}_{q}({\cal G},\Pi)$ only one pole is shielded, while
the second produces an unpleasant $p_{t}^{-1}$ behavior. However,
this does not lead to an infrared divergence of the cross-section, since
the same power $p_{t}^{-1}$
is implicitly present in the differential $dp_{t}^{2}$ on the l.h.s.

In Appendix~2. we consider in full the $s$-channel production of light
quarks from the process $gg\rightarrow q{\bar q}$.
The mathematical details behind
the above qualitative analysis can be found there. Here we shall discuss only
the main physical issues.

The higher powers of $p_t$ in the trace (\ref{eq:N13}) lead
to an increase in the differential cross-section at high $p_t$ over
the lowest order result. This is in compliance with the observation that
the first order terms bring more than a simple quantitative correction to the
lowest order. It is precisely the
emission of two back-to-back jets which makes possible the existence of a
high-$p_t$ particle in the final state.

\bigskip
\noindent {\bf \Large 7. Conclusion.}
\bigskip

We have considered new principles to compute the distributions
of quarks and gluons created in the first hard interaction of the two
heavy ions at high energies. We essentially employed an initial resummation of
the perturbation series for the probabilities \cite{Mak}.
It allowed us to describe
two different high energy processes, {\it viz.}, e-p scattering
and nuclear interactions, in the same terms, as two versions of
the same phenomenon -- deeply inelastic scattering of composite systems.

It is shown that these calculations can be performed without reference to
parton phenomenology. We have introduced the concept of a source as the
main subject of QCD evolution, and have shown that the equations which describe
the dynamics of the sources are independent of the type of
high-energy process, and independent of
the particular choice of the final interaction.

The additional benefit of the new approach is that it explicitly displays
the causal
structure of the QCD evolution equations, and their physical meaning as the
spectral analysis of the composite system as performed by the interaction which
results in the bare quark or gluon  production.
The evolution equations for the sources allow for a smooth transition between
the regimes described by the GLAP and BFKL equations.
The by-product of this
study is a new form of the fusion term in the GLR-type equation, which
might lead to the stronger low $x$ saturation of the sources than any terms
considered previously.

One of the most important results of this study is the new type of
perturbation expansion, which, unlike for the factorization technique,
does not lead to double counting of processes.   The diagrams already included
in the definition of the sources, and controlled in aggregate by the DIS data,
do not appear again in the higher orders of the new perturbative expansion.
The diagrams which do appear are free from initial state infrared
(collinear) singularities and do not require artificial cut-offs. The
leading parts of these diagrams do not depend upon the initial factorization
scale either. The price one pays for the efficiency of these
calculations, is that one requires the full $x$ and $Q^2$ dependence of
the sources (or structure functions) extracted from the data.

We are now in a position to calculate the single-particle distributions
of the  quarks and gluons after the first 0.1~fm  of the heavy ions collision.

\bigskip
{\bf \Large  Acknowledgements.}
\bigskip

I am grateful to L. McLerran, A.H. Mueller, E.Shuryak, J.Smith, A.Vainshtein
R.Venugopalan and G. Welke for many stimulating discussions.

\medskip
This work was supported by the U.S. Department of Energy under Contract
No. DE--FG02--94ER40831.

 \renewcommand{\theequation}{A1.\arabic{equation}}
\setcounter{equation}{0}
\bigskip
\bigskip
{\bf \Large  Appendix~1.~The full form of evolution equations.}
\bigskip

In the main body of the paper, we explicitly studied only
the part of evolution equations which eventually results in the GLAP
equations. The latter are also known as the leading logarithmic
approximation (LLA).

Corrections to Eqs.~(\ref{eq:G3}) and (\ref{eq:G4}) are of various
origin. Unfortunately, we cannot count these corrections
in the traditional way, which relies on the firm
hierarchy of twists in OPE-based calculations.
For the moment,  by ``next approximation'' we shall mean some
kind of structural expansion based on the complexity of the processes
taken into consideration. Surprisingly, it does not lead
far away from the commonly used scheme.

In what follows, we study two types  of corrections. First, we consider the
terms missing in the analysis of the simplest (by their structure) equations,
(\ref{eq:G3}) and (\ref{eq:G4}).  In the next two subsections we study the
low-$x$ region which results in BFKL equation ~\cite{BFKL} as the limit
of new evolution equations.
More complicated corrections lead to an equation resembling the
GLR equation ~\cite{GLR,MueQui}, but with some significant differences.

\bigskip
\bigskip
{\underline{\it A1.1.~Trivial corrections.}}
\bigskip

The simplest corrections arise from Eqs.~(\ref{eq:G3}) and (\ref{eq:G4}),
as a residue of  the original spinor and tensor form after LLA-terms
have been extracted.  To
make them more visible, let us split $\sigma_{1}(p)$ and
$w_{1}(p)$ into a leading ( $\sigma'_1$ and $w'_1 $) part,
corresponding to LLA,
and  subleading parts ($\sigma''_1$, $w''_1$,...)~:
\begin{equation}
 \sigma_{1}(p)=\sigma'_{1}(p)+\sigma''_{1}(p)+\sigma'''_{1}(p),\;\;\;\;
 w_{1}(p)=w'_{1}(p)+w''_{1}(p)+w'''_{1}(p) .
\label{eq:A1}
\end{equation}
The mathematical steps just follow those of the perturbation
calculations described in Section~3. Now it is only
a lengthy exercise to obtain the resulting equations.
Spinor components of the quark source may be expanded as follows:
\begin{eqnarray}
 \sigma'_{1}(p) = \int d^4 k  \Delta_{kp}
   {-p^2k^+ \over p^+}
\left[ P_{qq}({p^+ \over k^+}){k^+ \sigma_{1}(k)
 \over {\cal S}^{R}_{1}(k){\cal S}^{A}_{1}(k)  }
 + P_{qg}({p^+ \over k^+}) {k^+ w_{1}(k) \over
 {\cal W}^{R}_{1}(k){\cal W}^{A}_{1}(k) }  \right]
\label{eq:A2}
\end{eqnarray}
\begin{eqnarray}
 \sigma''_{1}(p) = \int d^4 k   \Delta_{kp} k^2
 \left[- P_{qq}({p^+ \over k^+}) {k^+ \sigma_{1}(k) \over
{\cal S}^{R}_{1}(k){\cal S}^{A}_{1}(k)  }
 + P_{qg}({p^+ \over k^+}) {k^+ w_{1}(k) \over
 {\cal W}^{R}_{1}(k){\cal W}^{A}_{1}(k) }   \right]
\label{eq:A3}
\end{eqnarray}
\begin{eqnarray}
 \sigma'''_{1}(p) =  \int d^4 k  \Delta_{kp} p^+
\left[ C_F {\sigma_{2}(k) \over {\cal S}^{R}_{2}(k){\cal S}^{A}_{2}(k)  }
+ (1-{p^+ \over k^+}) {k^2w_{2}(k) \over
{\cal W}^{R}_{2}(k){\cal W}^{A}_{2}(k)}\right]
\label{eq:A4}
\end{eqnarray}
\begin{eqnarray}
  \sigma_{2}(p) = \int  d^4 k \Delta_{kp} {k^+ \over p^+}
 \left[ C_F {k^+ \sigma_{1}(k) \over
{\cal S}^{R}_{1}(k){\cal S}^{A}_{1}(k)}
+ (1-{p^+ \over k^+}){k^+ w_{1}(k) \over
{\cal W}^{R}_{1}(k){\cal W}^{A}_{1}(k) }  \right]
\label{eq:A5}
\end{eqnarray}
where we have denoted
 \begin{eqnarray}
\Delta_{kp}={2 g_{r}^{2} \over (2\pi)^3 k^+}\delta^+[(k-p)^{2}]
={2 g_{r}^{2}\theta(k^0-p^0) \over (2\pi)^3 k^+} \delta[(p^+-k^+)(p^- -k^-)
 -({\bf p}_{t}-{\bf k}_{t})^{2}] \nonumber
\end{eqnarray}

The meaning of this decomposition becomes clear if we
multiply it by the kinematic factor $p^+$, and sandwich it between retarded
and advanced propagators (see Eq.~(\ref{eq:G14})).  The
joint left hand side of
Eqs.~(\ref{eq:A2})-~(\ref{eq:A4}) becomes the derivative with respect
to $Q^2$ of the quark structure function of  DIS.  The r.h.s. of
Eq.~(\ref{eq:A2}) then leads to the GLAP part of the evolution. The
factor $p^2$ in it is  responsible for the logarithmic behavior. Now it is
easy to see that r.h.s. of (\ref{eq:A3}) will result in a
term with $1/Q^2$ behavior. It simulates the next twist contribution,
even though the second twist is not included explicitly in the
density matrix as an effect of next order correlations.

The  r.h.s. of (\ref{eq:A4}) represents a secondary influence of the
longitudinal components of the spinor and gluon sources.
This components are defined
by  (\ref{eq:A5}) and (\ref{eq:A8}), and are of next order by a formal
count of the $\alpha_s$-powers. $\sigma_2$  directly contributes to the
longitudinal structure function
given by Eq. (\ref{eq:G16}). It also
appears in the Born term of the quark excitation process,
but we leave it aside for now since it is poorly controlled by the data.

All of the above comments apply equally to the components of the gluon source,
which can be decomposed following the same principle:
\begin{eqnarray}
 w'_{1}(p) = \int d^4 k  \Delta_{kp}
 {-p^2k^+ \over p^+}q
\left[ P_{gq}({p^+ \over k^+}) {k^+ \sigma_{1}(k) \over
 {\cal S}^{R}_{1}(k){\cal S}^{A}_{1}(k) }
  + P_{gg}({p^+ \over k^+}) {k^+ w_{1}(k) \over
 {\cal W}^{R}_{1}(k){\cal W}^{A}_{1}(k) }   \right]
\label{eq:A6}
\end{eqnarray}
\begin{eqnarray}
 w''_{1}(p)= \int  d^4 k  \Delta_{kp}  k^2
\left[ P_{gq}({p^+ \over k^+}) {k^+ \sigma_{1}(k) \over
 {\cal S}^{R}_{1}(k){\cal S}^{A}_{1}(k) }
  + P_{gg}({p^+ \over k^+}) {k^+ w_{1}(k)
\over {\cal W}^{R}_{1}(k){\cal W}^{A}_{1}(k) }     \right]
\label{eq:A7}
\end{eqnarray}
\begin{eqnarray}
 w'''_{1}(p) = \int  d^4 k   \Delta_{kp}
k^+ \left[ C_F  (1-{p^+ \over k^+}){\sigma_{2}(k)
 \over {\cal S}^{R}_{2}(k){\cal S}^{A}_{2}(k)  }
  -2N_c ({p^+ \over k^+}-{1\over 2})^2 { k^2 w_{2}(k)\over
 {\cal W}^{R}_{2}(k){\cal W}^{A}_{2}(k)}    \right]
\label{eq:A8}
\end{eqnarray}
\begin{eqnarray}
{w_{2}(p)\over p^2}= \int d^4 k  \Delta_{kp} {k^+ \over p^+}
[ -8 C_F n_f (1-{p^+ \over k^+})
{k^{+} \sigma_{1}(k)\over {\cal S}^{R}_{1}(k){\cal S}^{A}_{1}(k)}
+ 4 N_{c}{k^+ \over p^+} (1-{p^+ \over 2k^+})^{2}
{k^+ w_{1}(k) \over  {\cal W}^{R}_{1}(k){\cal W}^{A}_{1}(k) } ]
\label{eq:A9}
\end{eqnarray}

A preliminary examination of the additional terms in the evolution
equations reveals that they have
the same singular infrared behavior as GLAP equations,
and should be regularized
and renormalized. The way to do this is not yet clear, since conservation
of momentum seemingly fails to control all necessary subtractions.
A complete study of these equations is a separate subject. In this
paper, we have considered
the explicit form of the DIS structure functions as granted.

\bigskip
\bigskip
{\underline{\it A1.2.~The BFKL equation.}}
\bigskip

As has been shown in Sec.4, the new evolution equations for the
sources, even in their reduced form (\ref{eq:G3}) and (\ref{eq:G4}),
are practically the equations for the
derivatives of the structure functions, rather than for the DIS structure
functions themselves. The former were first introduced by Lipatov as
the ``unintegrated structure functions,'' and they relate to the latter
in the following way
\begin{equation}
f(x,p_{t}^{2}) = p_{t}^{2}{d ~x G(x,p_{t}^{2}) \over d p_{t}^{2}}
\label{eq:A21}\end{equation}
Thus, up to an insignificant normalization factor, we may write:
\begin{equation}
 \delta(p^-) f(x,p_{t}^{2})=
 {ix^2 ~p_{t}^{2} w^{01}_{1}(p)\over {\cal W}^{R}_{1}(p){\cal W}^{A}_{1}(p)}
\label{eq:A22}\end{equation}
In the limit of low $x$, the function $f(x,p_{t}^{2})$ was proven to obey
the so-called BFKL equation,
\begin{equation}
-x{\partial f(x,p_{t}^{2}) \over \partial x}=
{3\alpha_s(p_{t}^{2}) \over \pi} p_{t}^{2} \int_{0}^{\infty}
{d k_{t}^{2}\over k_{t}^{2}}
\left[{f(x,k_{t}^{2}) - f(x,p_{t}^{2}) \over |p_{t}^{2}- k_{t}^{2}|}
+{f(x,p_{t}^{2})\over \sqrt{4 k_{t}^{4} - p_{t}^{4}}}\right]
\label{eq:A23}\end{equation}
This equation was originally obtained by considering the amplitude
of the process
$2\rightarrow 2+n$ and summing the leading $log(1/x)$ terms \cite{BFKL}.

Evolution equations like (\ref{eq:G4}) were derived immediately
as integral equations.  At high $Q^2$ and not too low $x$, they
allow for simplifications which lead to the GLAP equations.
This means that the necessary resummation of the leading $log Q^2$
was done from the very beginning. Next,
our intention is to determine what simplifications should be done to
obtain the BFKL equation.
These simplifications are effectively equivalent to a
reduction in the number of diagrams already accounted for in the integral
equation.

In the notation of Eq.(\ref{eq:A21}), our equation (\ref{eq:G10}) for the
transverse gluon source reads
\begin{eqnarray}
 \delta(p^-) f(x,p_{t}^{2})=
 {\alpha_{s}(p_{t}^{2}) \over (2\pi^2} x p_{t}^{2}
\int_{x}^{1} {dy \over y}\int {d^2{\bf k}_{t}  \over k_{t}^{2} }
\delta[(p^+-k^+)p^- -({\bf p}_{t}-{\bf k}_{t})^{2}]
\left[-p^2 \over {\cal W}^{R}_{1}(p){\cal W}^{A}_{1}(p)\right]
 P_{gg}({x \over y})f(y,k_{t}^{2})
\label{eq:A24}
\end{eqnarray}
We have retained all terms that contribute to the LLA. We notice that
at real momenta the denominator $~{\cal W}^{R}_{1}(p){\cal W}^{A}_{1}(p)~$
is strictly positive both before and after any integration without
additional weight. Thus the full integral is not singular.
The next steps are as follows:

{\it Step 1.} ~~For $x\ll 1$, approximate the splitting kernel as
$ P_{gg}(x/y) \sim (y/x)$.

{\it Step 2.} Integrate both sides of the equation over $p^-$,
considering the retarded and advanced propagators in the
integrand on the r.h.s as bare. In this way the singular
infrared behavior of  the integrand is unshielded.
Integrate over the azimuth angle.

{\it Step 3.} Expand  the propagators on the l.h.s. up to the first order in
radiative corrections. Retain only the vacuum correlator and move it to the
r.h.s. in order to shield the singularity that resulted from
the approximation.

{\it Step 4.}  Differentiate both sides of the equation with respect
to $x$.

This procedure will result in the BFKL equation (\ref{eq:A23}) (up to the
insignificant last term in the integrand). The asymptotic behavior
of its solution is well known. Its exponential part,
\begin{eqnarray}
 f(x,p_{t}^{2}) \sim
\exp\bigg({-\ln^2 (p_{t}^{2}/{\bar p}_{t}^{2}) \over
2 \lambda'' \ln(x_0/x) }\bigg)
\label{eq:A25}
\end{eqnarray}
provides a decrease of $f(x,p_{t}^{2})$ at high $p_{t}^{2}$ that is
faster than any negative power of $p_{t}^{2}$.

We postpone any discussion of the accuracy of the above approximation,
and satisfy ourselves with the most important fact that
the evolution equations contain both GLAP and BFKL regimes of evolution
as the limits. Thus, we may hope to describe a
smooth transition between them. From a pragmatic point of view,
the above asymptotic behavior guarantees the convergence of
the integrals that appear in a calculation of the cross-sections of
quark and gluon production (see Sections 2 and 6).

\bigskip
\bigskip
{\underline{\it A1.3.~Gluon shadowing.}}
\bigskip

More complex corrections to the GLAP evolution correspond to the
replacement of
the $\#$-labelled correlators in Eqs.~(\ref{eq:G3}) and (\ref{eq:G4})
by correlators with sources, {\it viz.}, the second terms in
Eqs.~(\ref{eq:G1}) and (\ref{eq:G2}). This means that instead of the
final state correlators describing the emission, we include the initial
state correlators. This replacement accounts for
possible fusion of the partons.
Fusion is expected to be most important for gluons at low $x$. So,
only the purely gluonic component will be considered here.

Terms responsible for fusion of two gluon fields have the form
\begin{eqnarray}
\Delta_{fus}  \Pi^{ab,\mu\nu}_{01}(p)=ig_{r}^{2}
\int  {d^4 k d^4 q \over (2\pi)^4} \delta(k+q-p)
  V^{\mu\rho\sigma}_{acf}(k+q,-q,-k)\times \nonumber \\
\times\left[D_{ret}(k) \Pi_{10}(k) D_{adv}(k)\right]_{ff'}^{\sigma\beta}
 V^{\nu\lambda\beta}_{bc'f'}(-k-q,q,kk)
\left[D_{ret}(q) \Pi_{10}(q)
D_{adv}(q)\right]_{cc'}^{\rho\lambda}~ ~ ~.
\label{eq:A10}\end{eqnarray}
This equation is identical to Eq.~(\ref{eq:H15}) describing gluon fusion
in a nuclear collision, except that here both gluons are taken from the same
nucleus. We shall use the standard approximation, {\it i.e.},
bare tree propagators and bare
vertices. The longitudinal gluon function $w_2$ will be neglected also.
Routine calculations similar to those performed in Section~5 then yield:
\begin{eqnarray}
\Delta_{fus}G'(x,p_{t}^{2}) =
-{3\pi\alpha_0 \over 2 p_{t}^{4}} {(2\pi)^3 \over \pi R^2 } x^2
\int_{0}^{x}{dx_1\over x_1} \int_{0}^{x}{dx_2\over x_2}
\delta(x_1+x_2-x)
\int { dk_{t}^{2}dq_{t}^{2} \over 4S(k_t,q_t,p_t)}  \times \nonumber\\
\times \left[{x^2 \over x_1x_2}-{ x_1x_2 \over x^2}\right]^2
\left[{k_{t}^{2} \over x_1}+{q_{t}^{2} \over x_2}-{p_{t}^{2} \over x}\right]
G'(x_1,k_{t}^{2})G'(x_2,q_{t}^{2}), \hspace*{2cm}
\label{eq:A11}
\end{eqnarray}
where $G'(x,p_{t}^{2}) =dG'(x,p_{t}^{2})/dp_{t}^{2}$ and $S(k_t,q_t,p_t)$
is the area of a triangle with the sides $k_t$, $q_t$, and $p_t$.
The initial normalization factor $(V_{lab}P^+)^{-1}$, which was convenient
for calculation of cross-sections, has been replaced in
Eq.~(\ref{eq:A11}) by $(\pi R^2)^{-1}$, which corresponds to a
normalization per unit transverse area of a nucleus with radius $R$.

Eq.~(\ref{eq:A11}), by its structure, is very similar to the well known
Gribov-Levin-Ryskin (GLR) equation \cite{GLR}, and one derived later by Mueller
and Qui \cite{MueQui}.  It clearly  reveals the same tendency to saturate
the rate of the field source QCD-evolution  at low $x$. However, it
seems to have several differences. The most significant is that the
power of  $\alpha_s$  in  Eq.~(\ref{eq:A11}) is less than in Refs.~\cite{GLR}
and ~\cite{MueQui}.
The formal reason is simpler form of the
vertex of the ``three-ladder interaction''
that is prescribed for us by the general structure of the
evolution equations
(\ref{eq:G3}) and (\ref{eq:G4}). The other differences are of dynamic
origin and will be discussed elsewhere.

 \renewcommand{\theequation}{A2.\arabic{equation}}
\setcounter{equation}{0}
\bigskip
{\bf \Large  Appendix~2. Some estimates of the first order terms.}
\bigskip

\bigskip
\bigskip
{\underline{\it A2.1.~The ``box'' diagram.}}
\bigskip

Two box-type diagrams appear in the calculation of the quark
production cross-section, if we use the master formula (\ref{eq:N0}).
They are depicted in Fig.\ref{Fig5}, and the corresponding
analytic formula is~:
\begin{eqnarray}
2 p^0 { dN_{q}^{box} \over d{\bf p}d^4 x} ={-ig_{0}^{2} \over (2\pi)^3}
\int {d^4 k d^4 q \over (2\pi)^8 }
[{\rm Tr} \not p {\bf t}^a \gamma^{\mu} {\bf G}_{ret}(p-q)
{\bf t}^d \gamma^{\sigma}
{\bf G}^{\#}_{01}(p-k-q){\bf t}^c \gamma^{\rho} \times\nonumber\\
{\bf G}_{adv}(p-q){\bf t}^b \gamma^{\nu}
{\bf D}^{(A)dc}_{01,\sigma\rho}(k){\bf D}^{(B)ab}_{01,\nu\mu}(q)
+(A \leftrightarrow B)].\hspace*{2cm}
\label{eq:AA1}
\end{eqnarray}
Routine calculations result in the following expression for the differential
cross-section~:
\begin{eqnarray}
 {d\sigma^{(box)}_{q}({\cal G},{\cal G}) \over d p_{t}^{2} dy} =
-{2\pi \alpha_{0}^{2} \over 3 s p_{t}^{2}} \int_{0}^{1}
{dx_A dx_B \over x_{A}x_{B} }\theta (x_A-{p_{t}\over \sqrt{s}}e^{y})
\theta (x_B-{p_{t}\over \sqrt{s}}e^{-y}) \times  \nonumber   \\
\times \left[1-{p_{t}^{2}(3-x_Ax_B)\over sx_Ax_B}-
{4p_{t}^{4}\over (sx_Ax_B)^2} \right] G(x_A,Q_{0}^{2})G(x_A,Q_{0}^{2})
\delta [x_Ax_B-{p_{t}\over \sqrt{s}}({x_Ae^{-y} +x_B}e^{y})].
\label{eq:AA2}
\end{eqnarray}
The factor $p_{t}^{-2}$ appears
in the same way as in Eq.~(\ref{eq:H9}) for the
Born term. However, previously  it could be effectively absorbed into
the structure functions, which is not the case now. Indeed, Eq.~(\ref{eq:AA2})
has no additional integration over momenta which could be rescaled by $p_t$.
We expect at least a logarithmic divergence of total cross-section as a
result. This divergence
may be strengthened by the low-$x$ behavior of structure functions
because of the $p_t$ dependence of the
low limits of integration over $x_A$ and $x_B$ in Eq.~(\ref{eq:AA2}).

\bigskip
\bigskip
{\underline{\it A2.2.~Inclusive production of quarks in $s$-channel.}}
\bigskip

The analytic expression for the diagram (S2) of Fig.\ref{Fig3},
corresponding to the subprocess $gg\rightarrow q{\bar q}$, is~:
\begin{eqnarray}
2 p^0 { dN_q \over d{\bf p}d^4 x} ={-ig_{0}^{2} \over (2\pi)^3}
\int {d^4 k d^4 q \over (2\pi)^8 }
[{\rm Tr} \not p {\bf t}^a \gamma^{\mu}
{\bf G}^{\#}_{01}(p-k-q){\bf t}^b \gamma^{\nu}
{\bf D}^{(\#)aa'}_{ret,\mu\lambda}(k+q)
{\bf D}^{(\#)bb'}_{adv,\gamma\rho}(k+q)  \times\nonumber\\
\times\{ V^{\lambda\rho\sigma}_{acd}(-k-q,k,q)
{\bf D}^{(A)cc'}_{01,\rho\alpha}(k)
 V^{\gamma\alpha\beta}_{bc'd'}(k+q,-k,-q){\bf D}^{(B)dd'}_{01,\sigma\beta}(q)
+(A \leftrightarrow B) \}\hspace*{2cm}
\label{eq:AA3}
\end{eqnarray}
The most difficult problem here is to calculate the trace over
spinor and vector
indices. The complete expression is very unwieldy, and we will
therefore satisfy ourselves with
\begin{eqnarray}
{\rm Trace} \;=\;
12\times 16p_{t}^{2}sx_Ax_B[1+{\cal O}(p_t/\sqrt{s})], \nonumber
\end{eqnarray}
where the factor 12 comes from the color algebra.

As previously, we have three contributions to this process,
\begin{eqnarray}
 \sigma^{(2)}_{q}=\sigma^{(2)}_{q}({\cal G},{\cal G})+
\sigma^{(2)}_{q}({\cal G},\Pi)+
 \sigma^{(2)}_{q}(\Pi,\Pi).
\label{eq:AA4}
\end{eqnarray}
The first term corresponds to the factorization of the parton's
structure functions and the ``hard'' cross-section at a given scale,
\begin{eqnarray}
 {d\sigma^{(2)}_{q}({\cal G},{\cal G}) \over d p_{t}^{2} dy} =
{3 \alpha_{0}^{2}p_{t}^{2} \over 2 s^3} \int_{0}^{1}
{dx_A dx_B \over x_{A}^{2}x_{B}^{2} }
\theta (x_A-{p_{t}\over \sqrt{s}}e^{y})
\theta (x_B-{p_{t}\over \sqrt{s}}e^{-y})\times \nonumber \\
\times  G(x_A,Q_{0}^{2})G(x_B,Q_{0}^{2})
\delta [x_Ax_B-{p_{t}\over \sqrt{s}}(x_Ae^{-y}+x_Be^{y})].
\label{eq:AA5}
\end{eqnarray}
This term, explicitly dependent on the factorization scale, is kinematically
suppressed by three powers of $s^{-1}$.  Two of these come from the
kinematics of intermediate gluon, and the third from the phase volume which
is confined to a line in the $(x_A,x_B)$-plane.

The second term in Eq.~(\ref{eq:AA4})
describes the interaction of a parton with the source,
\begin{eqnarray}
 {d\sigma^{(2)}_{q}({{\cal G},\Pi}) \over d p_{t}^{2} dy} =
{3 \alpha_{0}^{2}p_{t}^{2} \over 16\pi^2} \int_{0}^{1}dx_A \int_{0}^{1}dx_B
\theta (x_A-{p_{t}\over \sqrt{s}}e^{y})
\theta (x_B-{p_{t}\over \sqrt{s}}e^{-y})\times \nonumber \\
\times\int_{|p_t-\xi|}^{p_t+\xi}{ k_t d k_t \over 4S(k_t,p_t,\xi)}
[{G(x_B,Q_{0}^{2})G'(x_A,k_{t}^{2})  \over
(s x_{A}^{2}x_{B}^{2}-k_{t}^{2})^2 } +(A \leftrightarrow B)]
\label{eq:AA6}
\end{eqnarray}
where we have introduced the notation,
\begin{eqnarray}
\xi^2=(k^+-p^+)(k^--p^-)=
s(x_A-{p_{t}\over \sqrt{s}}e^{y})(x_B-{p_{t}\over \sqrt{s}}e^{-y}). \nonumber
\end{eqnarray}
This second
term still keeps dependence on the factorization scale. Because of finite
virtuality of one of the incoming fields, the phase volume of the process is
larger, and consequently one of the powers $s^{-1}$ disappears.

The last term in Eq.~(\ref{eq:AA4}) accounts for the
interaction of two sources,
\begin{eqnarray}
 {d\sigma^{(2)}_{q}(\Pi,\Pi) \over d p_{t}^{2} dy} =
{3 \alpha_{0}^{2}p_{t}^{2} \over 32\pi^2} \int_{0}^{1}dx_A \int_{0}^{1}dx_B
\theta (x_A-{p_{t}\over \sqrt{s}}e^{y})
\theta (x_B-{p_{t}\over \sqrt{s}}e^{-y})\times \nonumber \\
\times\int_{|p_t-\xi|}^{p_t+\xi}{ l_t d l_t \over 4S(l_t,p_t,\xi)}
\int d^2 {\bf f} {G'(x_A,({\bf l}_{t}+{\bf f}_{t})^{2}/4)
G'(x_B,({\bf l}_{t}-{\bf f}_{t})^{2}/4) \over
(s x_{A}^{2}x_{B}^{2}-l_{t}^{2})^2 }~ ~ ~.
\label{eq:AA7}
\end{eqnarray}
Here, both incoming fields are virtual, and the $s$-channel propagators are
smoothed by the total transverse momentum of initial fields. The internal
integral over $d^2 {\bf f}$ represents the two-dimensional Fourier transform
of the product of the densities (in coordinate space) of two sources. Hence,
this term is proportional to the degree of geometrical overlap between the
colliding nuclei. It does not depend on a factorization scale, and is
expected to be the dominant term.

All three terms do not exhibit any
problems at low $p_t$, but a reliable knowledge
of the low $x$ behavior of the sources may be important for quantitative
computations.  The presence
of a factor $p_{t}^{2}$, which is purely kinematic in its
origin, guarantees that at high $p_{t}$ the first order
differential cross-section will be larger than the
Born cross-section in this region.

% now the references. delete or change fake bibitem. delete next three
%   lines and directly read in your .bbl file if you use bibtex.

% figures follow here
%
% Here is an example of the general form of a figure:
% Fill in the caption in the braces of the \caption{} command. Put the label
% that you will use with \ref{} command in the braces of the \label{} command.
%
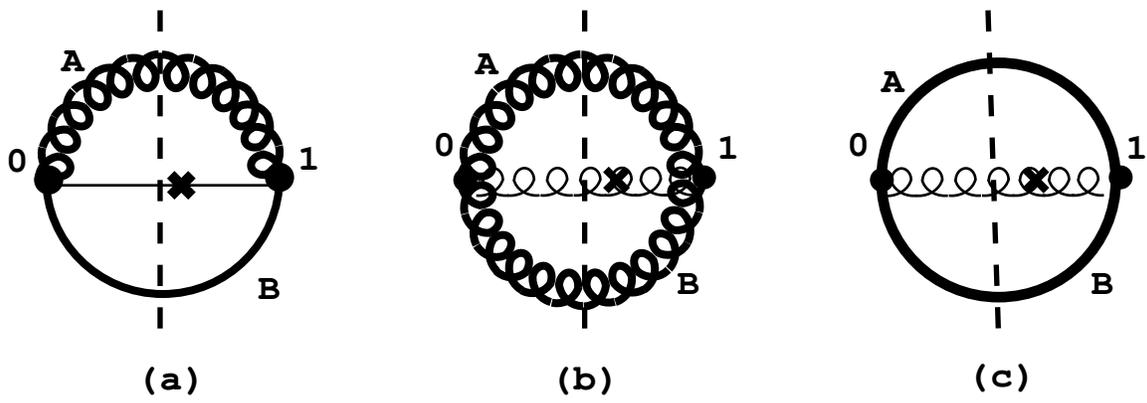
\begin{figure}
\caption{Born diagrams for the one quark and one gluon
production. The bold cross labels the line corresponding to the
``detected'' particle with momentum $p$. The dashed line crosses
the Greenians representing densities of the initial (bold) or final
(thin) states. Numbers near the vertices indicate the type of
ordering in the Greenians.
The processes: (a) $qg\rightarrow q$; (b) $gg\rightarrow g$;
(a) $q{\bar q}\rightarrow g$. }
\label{Fig1}
\end{figure}
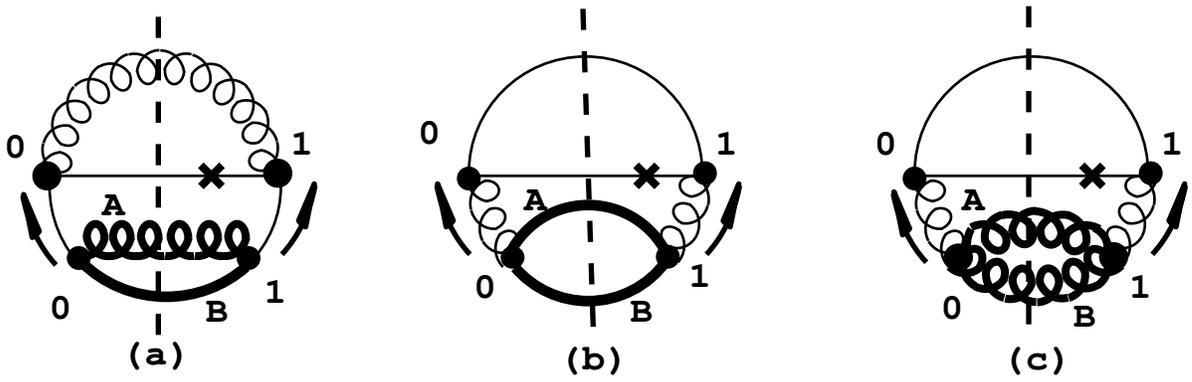
\begin{figure}
\caption{Self-energy-type  first order
diagrams for the one quark
production.  Notation the same as in Fig.1. Arrows label
the retarded and advanced propagators and show the latest
time. }
\label{Fig2}
\end{figure}
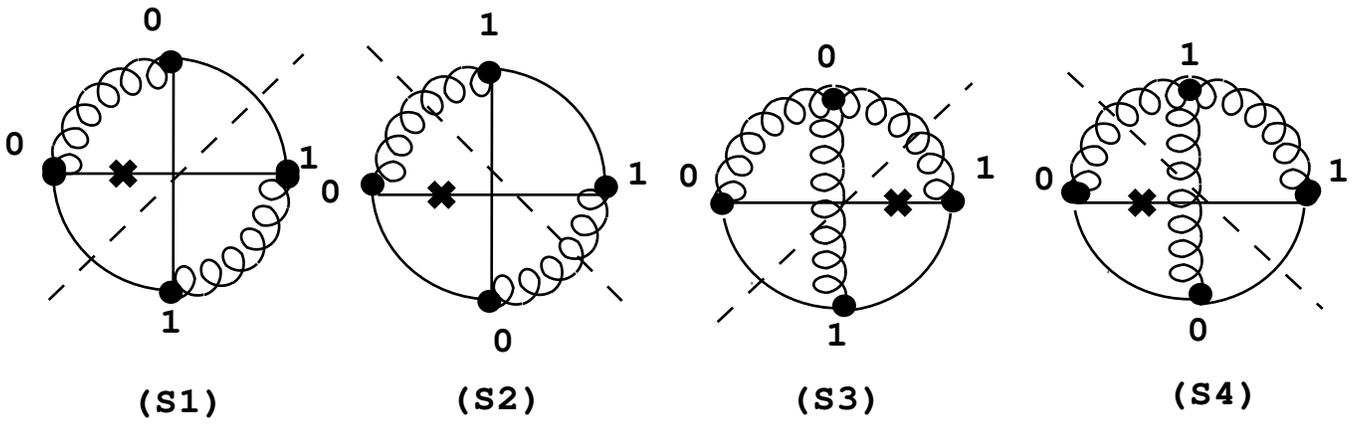
\begin{figure}
\caption{Four topologies of the vertex-type  diagrams for the one
quark production.}
\label{Fig3}
\end{figure}
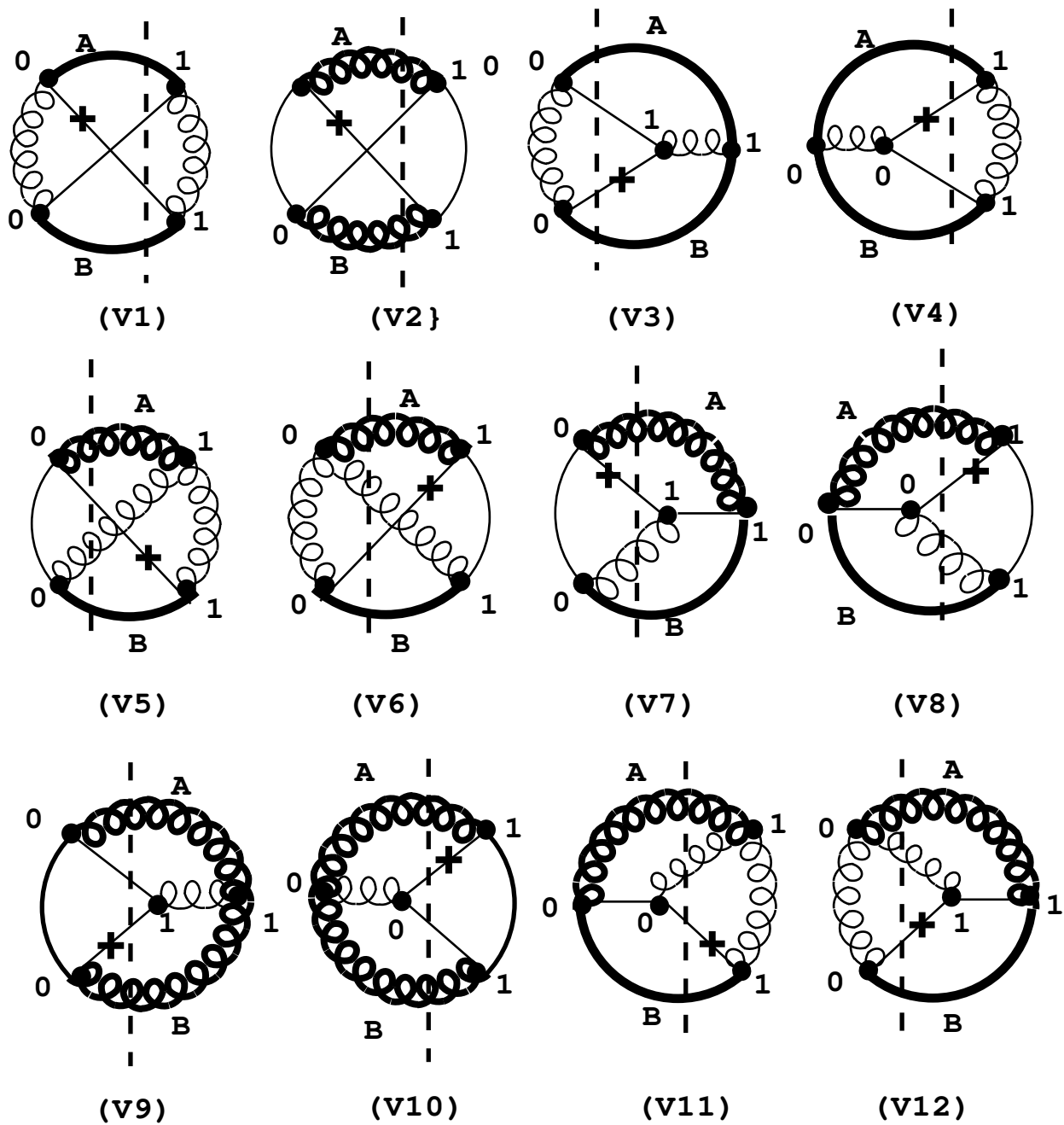
\begin{figure}
\caption{Twelve vertex-type  diagrams of for the one quark
production in the first order.}
\label{Fig4}
\end{figure}
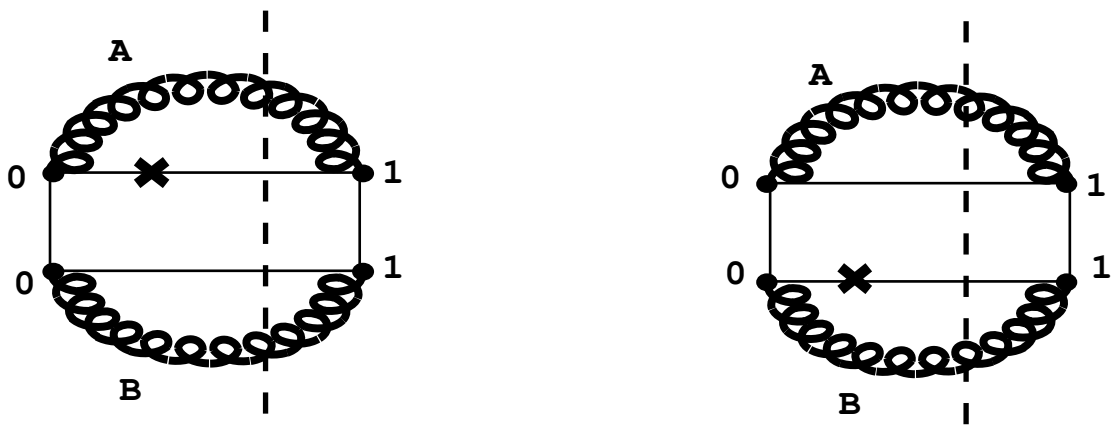
\begin{figure}
\caption{Two infrared-divergent ``box'' diagrams which are not
generated by our perturbation theory.}
\label{Fig5}
\end{figure}

\end{document}